\documentclass[11pt]{article}
\usepackage{epsfig, amssymb}
\newcommand{\be}{\begin{equation}}
\newcommand{\ee}{\end{equation}}
\newcommand{\bea}{\begin{eqnarray}}
\newcommand{\eea}{\end{eqnarray}}
\newcommand{\bA}{\begin{array}}
\newcommand{\eA}{\end{array}}
\newcommand{\bc}{\begin{center}}
\newcommand{\ec}{\end{center}}
\newcommand{\al}{\alpha}

\newcommand{\ra}{\rightarrow}
\newcommand{\del}{\partial}
\newcommand{\ie}{{\it i.e.}}
\newcommand{\eg}{{\it e.g.}}

\newcommand{\Nf}{${\cal N}{=}4$}

\newcommand{\vx}{{\vec x}}

\newcommand{\cO}{{\cal O}}
\newcommand{\lA}{\langle}
\newcommand{\rA}{\rangle}

\begin{document}

\begin{titlepage}

\bc

\hfill  {Duke-CGTP-02-09} \\
\hfill  {\tt hep-th/0211110} \\
        [22mm]

{\Huge Blocking up D-branes : \\ \vspace{2mm} Matrix Renormalization ?}\\
\vspace{6mm}

{\Large K.~Narayan} \\
\vspace{3mm}
{\small \it Center for Geometry and Theoretical Physics, \\}
{\small \it Duke University, \\}
{\small \it Durham, NC 27708.\\}
\vspace{1mm}
{\small Email : narayan@cgtp.duke.edu}\\

\ec
\medskip
\vspace{22mm}

\begin{abstract}
Drawing analogies with block spin techniques used to study continuum
limits in critical phenomena, we attempt to block up D-branes by
averaging over near neighbour elements of their (in general
noncommuting) matrix coordinates, \ie\ in a low energy description. We
show that various D-brane (noncommutative) geometries arising in
string theory appear to behave sensibly under blocking up, given
certain key assumptions in particular involving gauge invariance. In
particular, the (gauge-fixed) noncommutative plane, fuzzy sphere and
torus exhibit a self-similar structure under blocking up, if some
``counterterm'' matrices are added to the resulting block-algebras. 
Applying these techniques to matrix representations of more general 
D-brane configurations, we find that blocking up averages over 
far-off-diagonal matrix elements and brings them in towards the diagonal,
so that the matrices become ``less off-diagonal'' under this process. 
We describe heuristic scaling relations for the matrix elements under 
this process. Further, we show that blocking up does not appear to 
exhibit any ``chaotic'' behaviour, suggesting that there is sensible 
physics underlying such a matrix coarse-graining. We also discuss 
briefly interrelations of these ideas with B-fields and noncommutativity.
\end{abstract}

\end{titlepage}

\newpage 
\begin{tableofcontents}
\end{tableofcontents}

\vspace{5mm}

\section{The basic ideas and motivations}

Noncommutative geometry arises in various situations in string theory
\cite{NCreview}. In particular, the low energy dynamics of D-branes
stems from the low lying fluctuations of open strings stretched between
them. This gives rise to nonabelian gauge theories with charged scalar
fields that describe transverse fluctuations of the D-brane worldvolumes 
as well as other matter fields. Various D-brane configurations are then
described in terms of a set of commutation relations among the scalars 
(regarded here as matrices)
\be \label{branebgnd}
[X^i, X^j] = i \theta^{ij}(X)
\ee
$\theta^{ij}(X)$ is an antisymmetric 2-form that in general depends on the
coordinate variables $X^i$ (we restrict attention here to spatial
noncommutativity, \ie\ $\theta^{0i}=0$). When the $X^i$ commute with each
other, \ie\ $\theta^{ij}=0$, the eigenvalues of the matrices $X^i$ can be
interpreted as spacetime coordinates describing the positions of the 
D-branes. However, in general, the coordinate variables $X^i$ do not 
commute so that ``points'' on the space (\ref{branebgnd}) cannot be 
specified with arbitrary accuracy. This gives rise to a geometry defined
by noncommuting matrices \cite{witten9510}. In \eg, the construction 
\cite{banksseibergshenker} of BPS states in Matrix theory \cite{BFSS} 
(see also \cite{IKKT}), or the Myers dielectric effect \cite{myers99}, 
the space (\ref{branebgnd}) can be interpreted as a higher dimensional 
brane.

Whenever the coordinate variables $X^i$ do not commute, there are operator
ordering ambiguities in defining observables as functions of these
noncommuting coordinates. On the other hand, there are no such ambiguities
in defining the corresponding function in the commutative limiting 
geometry. In general, several classes of functions on the noncommutative
geometry collapse onto the same function in the commutative limit. Thus 
various noncommuting observables on a given noncommutative geometry reduce
to the same commutative observable. In this sense, the commutative 
description is a smooth coarse-grained approximation to the underlying 
microscopic D-brane description -- the noncommutative description 
contains more information than the commutative approximation\footnote{It
is useful to bear in mind the work of \cite{witten86}, \cite{cds9711},
\cite{schom9903}, \cite{sw9908} and \cite{berenleigh0005} on 
noncommutativity arising from the open string sector, the version of 
noncommutative algebraic geometry in \cite{berenleigh0005}, various 
gauge/gravity dualities \cite{malda9905}, as well as relations thereof
to tachyon condensation stemming from the ideas of Sen \cite{sen}.}.

Thinking thus of the commutative limit as a coarse-grained or averaged
approximation to the underlying noncommutative background is reminiscent 
of, \eg\ thinking of a continuum field theory as a coarse-grained
smooth approximation to a spin system near a critical point, or a
lattice gauge theory -- the underlying lattice structure is the
microscopic description but near the critical point, there are long
range correlations and one can block up spins to define effective
lattice descriptions iteratively and thereby move towards the
continuum limit \cite{blockRG}. Thus it is tempting to ask if one can
apply similar ideas and ``block up D-branes'' by averaging over
nearest neighbour open string modes at the level of matrix variables,
thereby moving towards the commutative limit. In what follows, we show
that these ideas do indeed lead to interesting physics.

It is important to note that since gauge transformations cannot undo
the noncommutativity, this means that off-diagonal open string modes
necessarily have nonzero vevs in such a configuration. A specific
solution to (\ref{branebgnd}) can be expressed in terms of a set of
representation matrices with necessarily nonzero off-diagonal matrix
elements.  Since the off-diagonal modes in general have nonzero vevs,
it is plausible to think of the resulting noncommutative geometry as a
condensate of the underlying D-brane (and therefore open string)
degrees of freedom that build up the geometry, as outlined
previously. Thinking of the noncommutative geometry as akin to a
condensed matter system near a critical point, it is tempting to think
of the open string modes as developing ``long range correlations'',
leading to a condensation of some of the modes. For example, as a
lattice spin system approaches its critical point, long range
correlations develop and patches of correlated spins emerge. One can
then block up the underlying microscopic spin degrees of freedom to
construct effective block spin degrees of freedom, as in the Kadanoff
scaling picture \cite{blockRG}. Holding the block spins fixed and
averaging over the microscopic spins yields a new theory describing
the interactions of these block spins, with new parameters that have
been scaled appropriately to reflect the block transformation.

We would like to take these ideas almost literally over to D-brane
matrix variables, where we then proceed to average over ``nearest
neighbour'' open string modes to generate effective D-brane
geometries.  We define block D-branes at the level of the effective
low energy description by averaging over nearest neighbour matrix
elements to define new matrix representations of (\ref{branebgnd}) and
thus a blocked-up or averaged background\footnote{Note that
\cite{vaidya} uses possibly related renormalization methods to analyse
scalar field theories on the fuzzy sphere. We also mention in passing,
\eg\ \cite{brezinzinnjustin92}, in the context of matrix models of 2D
quantum gravity.}. The new matrix representations generated under
blocking up have ``renormalized'' matrix elements. For the
noncommutative 2-plane, we hold the distances between points in the
commutative limiting 2-plane fixed as we block up -- this preserves
the translation isometries of the 2-plane. Iterating the matrix
blocking procedure moves towards progressively more coarse-grained
descriptions. Blocking up in this fashion turns out to shrink the
off-diagonal modes thus reducing the noncommutativity. In general,
blocking up brings far-off-diagonal modes in, towards the
diagonal. The matrices become ``less off-diagonal'' in the sense that
if initially they satisfy the scaling relation $\ X_{ab}\sim X_{aa}\
q^{|b-a|}\ $ for some $q<1$, then the resulting block-matrices satisfy
a similar relation with ${\tilde q} \sim q^2 ((1+q)/2q)^{1/(b-a)}$,
showing that far-off-diagonal modes scale away fast relative to the
near-off-diagonal ones. We find nontrivial conditions on the scalar
matrix elements for blocking up to move towards a commutative
limit. Furthermore, we find that blocking up does not seem to exhibit
any ``chaotic'' behaviour, thus indicating that there is sensible 
physics underlying such a matrix coarse-graining.

In this work, we make the nontrivial assumption that gauge invariance has
been fixed to yield a set of matrices that are best suited to recovering
a commutative limit. This turns out to be equivalent to assuming that
the matrix elements are ordered in specific ways. Physically this 
essentially picks a convenient gauge in which D-brane locations at the
level of matrix variables are in some sense identified with D-brane 
locations in physical space. The interesting (and more difficult) 
question of what the gauge-invariant physical matrix coarse-graining is, 
that this work is a gauge-fixed version of, will be left for the future.

Roughly speaking, one expects a smooth space only if one pixellates
the space using small enough pixels, in other words, a large number of
``points'' or D-branes. Thus in general, one expects that any such
block brane techniques would only make sense in a large $N$ limit (see
for \eg\ \cite{alekreckschom} for a worldsheet description of D-branes
on a fuzzy sphere). It is also useful to recall the ideas of
deconstruction \cite{decons} where certain quiver theories (in some
regions of the moduli space) can be thought of as developing extra
dimensions which are the quiver space, so that in some sense D-brane
positions are identified with lattice sites in the deconstructed
dimensions. At low energies, the effective theory does not see the
fact that the extra dimensions are really lattice-like and a continuum
description emerges\footnote{\cite{knrp0309} studies quiver theories
under a similar coarse-graining.}.

Some words on the organization of this work : Sec.~2 lays out the basic
definitions while Sec.~3 analyzes the noncommutative plane detailing the
consequences of blocking up and the commutative limit, outlining possible 
interrelations with noncommutative field theories and finally the role 
of gauge invariance. Sec.~4 studies other nonabelian geometries under
blocking up. Sec.~5 studies some general properties of matrices and 
noncommutative algebras under blocking up. Sec.~6 describes some 
conclusions and speculations. Finally, appendix~A describes the 
commutative case and appendix~B describes an alternative way to block up
D-branes and matrices.
\begin{figure}
\bc
\epsfig{file=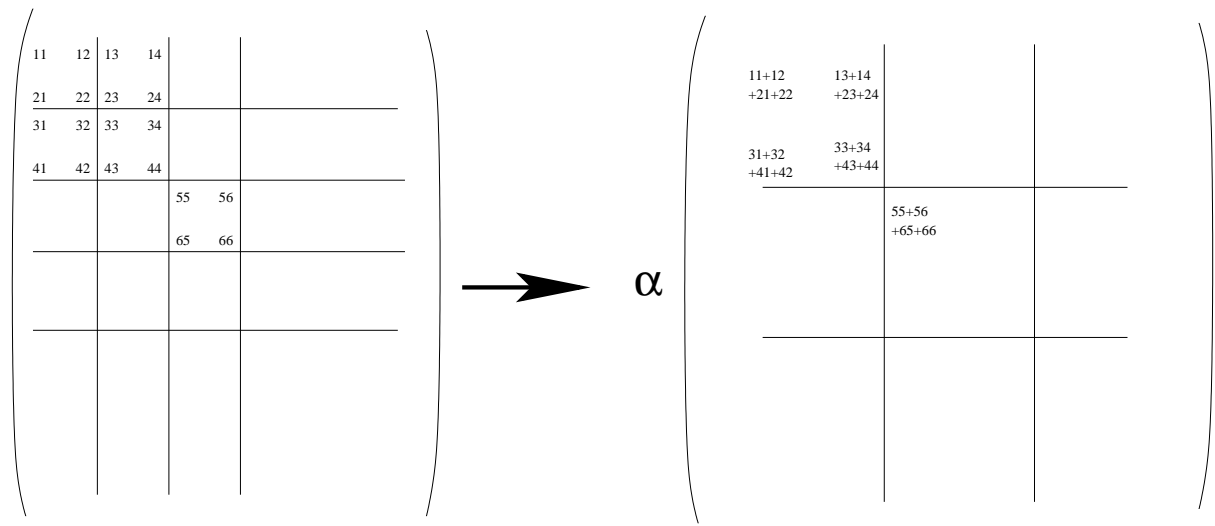, width=16cm}
\caption{Blocking up elements of the original matrix to get a new matrix
($2\times2$ blocks shown).}
\label{fig1}
\ec
\end{figure}

\section{Blocking up D-branes and matrices : definitions}

The scalars $X^i$ are hermitian matrices in the adjoint representation
of the gauge group, which we take for concreteness to be $U(N)$ for
the present. $N$ is assumed to be essentially infinite. Thus the $X^i$ 
satisfy
\be
X^i_{a,b} = {X^i}^{\dag}_{a,b} = {X^i}^*_{b,a}, 
\qquad \qquad \qquad a,b=1 \ldots N
\ee
Define a blocked up matrix as
\be \label{defnblock}
{\tilde X}^i_{a,b} = \alpha (X^i_{2a-1,2b-1} + X^i_{2a,2b-1} 
+ X^i_{2a-1,2b} + X^i_{2a,2b}),
\qquad \qquad a,b=1 \ldots {N\over 2}
\ee
In other words, define a matrix entry in the blocked up matrix by
averaging over the nearest neighbour matrix entries in $2\times 2$ blocks
of the original matrix. This is easier to see pictorially (see 
figure~\ref{fig1}).

The overall factor $\alpha$ has been put in as the normalization constant
for uniform averaging. We expect that a uniform normalization constant 
means that we have averaged uniformly all over the matrix\footnote{In 
general, we could imagine nonuniform averaging over the matrix. Hazarding
a guess, we expect that nonuniformities imply anisotropies in the matrix
space and hence perhaps anisotropies in spacetime.}. If we demand that the 
identity matrix is preserved under blocking up, \ie\ ${\bf 1} \ra {\bf 1}$,
it is clear that for $2\times 2$ blocks, we must fix $\ \al={1\over 2}$.
In general, we note that the averaging constant can be different for 
different matrices, \ie\ $\al=\al^i$.

It is clear from the definition of blocking (\ref{defnblock}) that this
preserves hermiticity -- the ${\tilde X}^i$s are also hermitian, so that 
they are also interpretable as scalars in the adjoint representation of a 
gauge group. Further, it is also clear that blocking up is (pictorially)
symmetric about the diagonal. One would further guess that blocking up in
this fashion would average far-off-diagonal modes and bring them in, 
towards the diagonal. This turns out to be almost (but not quite) true, 
as we shall see in section 4.

Finally note the following analogy with spin systems near a critical
point : defining block spins by averaging over $k\times k$ blocks instead
of $2\times 2$ blocks does not change the physics, yielding the same
Wilson renormalization group equations. This independence of the block 
size reflects long range correlations in the system. We expect that if
the noncommutative geometry at hand is truly a ``condensate'' exhibiting
long range correlations, defining block D-branes by averaging over blocks
of varying sizes should not change the physics. In the examples that 
follow, we shall see that this is indeed the case. Note that to preserve
the identity matrix under $k\times k$ blocks, \ie\ ${\bf 1} \ra {\bf 1}$,
we must fix $\ \al={1\over k}$.

A priori, it is not at all obvious that blocking up matrices in this
fashion is a sensible thing to do and if it will lead to any tractable
physics. We will see in what follows that it in fact does, if certain 
key assumptions are made -- in particular, we fix gauge invariance and 
identify D-brane locations at the level of matrix variables with D-brane 
locations in physical space. This gauge fixing enables remarkable 
simplifications and leads to recognizable structures which can then be 
interpreted in a relatively straighforward fashion. This nearest 
neighbour matrix element averaging is analogous to real space 
renormalization group methods in critical phenomena and so is perhaps 
expected to have all the weaknesses of those techniques. A more elegant 
renormalization group formulation of such a decimation analogous to 
integrating out thin momentum shells is lacking at this time.

\section{$\theta^{ij} = constant$ : the noncommutative plane}

The algebra representing the noncommutative or quantum 
plane\footnote{Multiple planes can be analyzed by applying the following
techniques independently in each plane (see later section).} is
\be \label{ncplanealg}
[X^1, X^2] = i \theta^{12} {\bf 1}
\ee
Such algebras arise in the construction \cite{banksseibergshenker} of 
infinite BPS branes in Matrix theory \cite{BFSS} where the brane 
worldvolume coordinates $X^1,X^2$ are noncommutative while the rest
commute. It is also the algebra of open string vertex operators on a 
D-brane boundary, in the presence of a constant $B_{NSNS}$-field 
\cite{NCreview}. $\theta^{12}$ is inversely proportional to 
$B_{NSNS}$ in the Seiberg-Witten limit \cite{sw9908}. This algebra is
preserved under translations by two independent matrices proportional to
the identity matrix
\be\label{ncpoints}
(X^1_0,X^2_0) \ra (X^1_0,X^2_0) + (x^1 {\bf 1}, x^2 {\bf 1}),
\ee
where $(X^1,X^2)$ is really a ``matrix point'' defined with a resolution
size -- thus the algebra defines a two-plane. Using a representation in 
terms of harmonic oscillator creation/annihilation operators satisfying 
$\ [a, a^{\dag}] = {\bf 1},\ $ we can write the $X^i$ as
\be\label{NCsoln}
X^1 = x^1 {\bf 1} + \sqrt{\theta^{12}} {1\over \sqrt{2}}(a + a^{\dag}), 
\qquad 
X^2 = x^2 {\bf 1} + \sqrt{\theta^{12}} {i\over \sqrt{2}}(-a + a^{\dag})
\ee
where $(x^1,x^2)$ label a commutative point on the 2-plane. This is 
thus a decomposition into a ``center-of-mass'' piece and an off-diagonal
piece independent of the location. Thus, in this representation, it 
seems reasonable to interpret the noncommutative plane as a fuzzy 
resolution of the commutative plane -- at each point $(x^1,x^2)$ of the
commutative limit, we have a ``fuzzy blowup'' in terms of noncommuting 
matrices defining a characteristic pixellation size $\theta^{12}$.

\subsection{Blocking up}

We would like to preserve the translation isometries (\ref{ncpoints}) of
the 2-plane under blocking up. These translations are generated by the
identity matrix ${\bf 1}$. Further the commutative ``point'' is 
$(x^1{\bf 1},x^2{\bf 1})$ which we hold fixed under blocking up. This
requires that we set the averaging constant $\al={1\over 2}$.\\
Using the convenient basis of energy eigenstates of the harmonic 
oscillator, we have for the representation matrices of $a, a^{\dag}$,
\bea
&& \qquad \qquad \qquad N|n\rA = a^{\dag} a |n\rA = n|n\rA, \nonumber\\
{} && \lA n'|a|n\rA = \sqrt{n} \ \lA n'|n-1\rA = \sqrt{n} \ \delta_{n',n-1}, 
\qquad \qquad \lA n'|a^{\dag}|n\rA = \sqrt{n+1} \ \delta_{n',n+1}
\eea
Let us order the states as 
\be\label{basisorder}
|0\rA = (1,0,0,\ldots)^T, \ \qquad 
|1\rA = (0,1,0,\ldots)^T, \ \qquad 
|2\rA = (0,0,1,0,\ldots)^T, \ \ldots \nonumber\\
\ee
With this basis, we get for the representation matrices of $X^1$
\bea
X^1 = x^1{\bf 1} + \sqrt{\theta^{12}\over 2}
\left( \bA{cccccccc} 0 & 1 & 0 & 0 & & & & \ldots \\ 
1 & 0 & \sqrt{2} & 0 & & & & \ldots \\ 
0 & \sqrt{2} & 0 & \sqrt{3} & 0 & & & \ldots \\
0 & 0 & \sqrt{3} & 0 & \sqrt{4} & 0 & & \ldots \\
& & 0 & \sqrt{4} & 0 & \sqrt{5} & 0 & \\
& & & & \sqrt{5} & 0 & \sqrt{6} &  \\
& & & & . & & &  \\
& & & & . & & &  \\
\eA \right)
\eea
the representation matrix for $X^2$ being easily constructed.
Blocking up this matrix in the way defined in the previous section gives
\bea
X^1&\ra& x^1{\bf 1} + {1\over 2} \sqrt{\theta^{12}\over 2}
\left( \bA{cccccc} 2 & \sqrt{2} & 0 & 0 & & \ldots \\ 
\sqrt{2} & 2\sqrt{3} & \sqrt{4} & 0 & & \ldots \\ 
0 & \sqrt{4} & 2\sqrt{5} & \sqrt{6} & 0 & \ldots \\
0 & 0 & \sqrt{6} & 2\sqrt{7} & \sqrt{8} & \ldots \\
& & 0 & \sqrt{8} & 2\sqrt{9} & \ldots \\
& & & & 0 &  \\
& & & . & &  \\
& & & . & &  \\
\eA \right) \nonumber\\
{}&=&x^1{\bf 1}+\sqrt{\theta^{12}\over 2}
\left( \bA{cccccc} 1 & 0 & 0 & 0 & & \ldots \\ 
0 & \sqrt{3} & 0 & 0 & & \ldots \\ 
0 & 0 & \sqrt{5} & 0 & & \ldots \\
0 & 0 & 0 & \sqrt{7} & & \ldots \\
& & & . & &  \\
& & & . & &  \\
& & & . & &  \\
& & & . & &  \\
\eA \right) + {\sqrt{2}\over 2} \sqrt{\theta^{12}\over 2}
\left( \bA{cccccc} 0 & 1 & 0 & 0 & & \ldots \\ 
1 & 0 & \sqrt{2} & 0 & & \ldots \\ 
0 & \sqrt{2} & 0 & \sqrt{3} & & \ldots \\
0 & 0 & \sqrt{3} & 0 & & \ldots \\
& & 0 & \sqrt{4} & & \\
& & & . & &  \\
& & & . & &  \\
& & & . & &  \\
\eA \right)\nonumber\\
\eea
In the last line, we have separated the diagonal matrix from the 
off-diagonal one. Blocking up works in a very similar fashion for $X^2$,
with the diagonal matrix generated having zero entries\footnote{In 
terms of energy eigenstates, we have essentially blocked up nearest
neighbour states as
\be
|n\rA^{block} = {1\over \sqrt{2}} \biggl(\ |2n\rA + \ |2n+1\rA \ \biggr), 
\qquad \qquad \qquad n=0,1,2\ldots
\ee
The normalization is fixed to agree with our previous matrix element 
blocking up, \ie, $a'_{mn} = \lA m'|a|n'\rA = a^{\rm blocked-up}$,
using (\ref{defnblock}). The new blocked states are not eigenstates of 
the original harmonic oscillator number operator (and therefore the 
oscillator Hamiltonian).}. 
Thus under blocking up,
\bea\label{newnoncoords}
{\tilde X}^1 &=& x^1 {\bf 1} + \sqrt{\theta^{12}\over 2} D^1 + 
{1\over \sqrt{2}} \sqrt{\theta_{12}\over 2} \ (a+a^{\dag}), \nonumber\\
{\tilde X}^2 &=& x^2 {\bf 1} + \sqrt{\theta^{12}\over 2} D^2 + 
{1\over \sqrt{2}} \sqrt{\theta^{12}\over 2} \ i (-a + a^{\dag}),
\eea
where the $D^i$ are the diagonal matrices that were generated ($D^2=0$).
Recall that we have held the commutative coordinates 
$(x^1{\bf 1},x^2{\bf 1})$ and the translation isometries fixed (by
setting $\al={1\over 2}$) -- thus distances between points 
$(x^1{\bf 1},x^2{\bf 1})$ are also held fixed as we block up. The 
algebra of the new coordinates is
\be \label{newalg}
[{\tilde X}^1, {\tilde X}^2]
= i {\theta^{12} \over 2} {\bf 1} + \Delta \theta^{12}
\ee
where
\be
\Delta \theta^{12} =
{\sqrt{2}\over 2} \theta^{12} \biggl([D^1,{1\over \sqrt{2}}(-a+a^{\dag})]
\biggr)
\ee
are nonzero since the diagonal matrices $D^i$ do not in general commute 
with an arbitrary matrix (the terms involving $D^2=0$ vanish). We will 
suggestively call the $D^i$ ``counterterm'' matrices. Thus the algebra
of the new noncommutative coordinates is ``self-similar'' in form
under blocking up, modulo counterterms involving the diagonal matrices
generated under blocking up. Notice further that the new $\theta^{12}$
is scaled down by a factor of $2$. Continuing this blocking up
procedure, after $n$ iterations, we have
\bea\label{newnoncoordsn}
X^{1 {(n)}} &=& x^1 {\bf 1} + \sqrt{\theta^{12}\over 2} D^{1 {(n)}} + 
{1\over 2^{n/2}} \sqrt{\theta^{12}\over 2} \ (a+a^{\dag})
\equiv x^1 {\bf 1} + \sqrt{\theta^{12}\over 2} D^{1 {(n)}} + \lambda 
(a+a^{\dag}) \nonumber\\
X^{2 {(n)}} &=& x^2 {\bf 1} + \sqrt{\theta^{12}\over 2} D^{2 {(n)}} + 
{1\over 2^{n/2}} \sqrt{\theta^{12}\over 2} \ i (-a+a^{\dag})
\equiv x^2 {\bf 1} + i \lambda (-a+a^{\dag})
\eea
where $\ \lambda\sim{1\over 2^{n/2}}\ $ is the scaling-down factor and 
the $kk$th-diagonal element of the $D^{1 {(n)}}$ is 
\be
{D^{1 {(n)}}}_{kk} = {1\over 2^{n-1}} \biggl( \sqrt{2^n.(k-1)+1} + 
\sqrt{2^n.(k-1)+2} + \ldots + \sqrt{2^n.(k)-1} \biggr)
\ee
There are $2^n.(k) - 2^n.(k-1) - 1 = 2^n - 1$ terms inside the bracket,
so that the diagonal elements ${D^{1 {(n)}}}_{kk}$ do not shrink to zero 
along with the off-diagonal elements. Indeed
\be
lim_{n\ra \infty} \ {D^{1 {(n)}}}_{kk} \sim 2^{(n+2)/2} \sqrt{k} \neq 0
\ee
so that blocking up does not scale down the diagonal elements of $D^1$, 
while scaling down the off-diagonal ones ($D^2 = 0$). The algebra of the
new coordinates after $n$ iterations is
\be \label{algitn}
[{\tilde X}^1, {\tilde X}^2]
= i {\theta^{12} \over 2^n} {\bf 1} + \Delta \theta^{12}
\ee
where the counterterms are
\be
\Delta \theta^{12} = 
{1\over 2^{n/2}} \theta^{12} \biggl([D^{1 {(n)}},
{1\over \sqrt{2}}(-a+a^{\dag})]
\biggr)
\ee
We see from (\ref{algitn}) that $\theta^{12}$ shrinks exponentially with
each iteration while the commutative coordinates are held fixed. Thus the
limit of the blocking up transformation is $\theta^{12} = 0$.

Averaging over $3\times 3$ blocks, say, or larger blocks turns out to be
equivalent to averaging over $2\times 2$ blocks. Indeed, averaging 
uniformly over $k\times k$ blocks, we see a periodic repetition of 
off-diagonal blocks involving factors of $\sqrt{k}$ which thus exhibits
a self-similar structure like the one above with $2\times 2$ blocks. 
Thus we find that the new noncommutative pieces of the coordinates scale
after $n$ iterations as $a \ra \biggl({\sqrt{k}\over k}\biggr)^n a, \ \
a^{\dag} \ra \biggl({\sqrt{k}\over k}\biggr)^n a^{\dag}$ after 
subtracting the diagonal matrices, so that $\theta^{12}$ shrinks as
\be
\theta^{12} \ra \theta^{12} \ {1\over k^n}
\ee
thus shrinking noncommutativity.
This is again reassuring, since if indeed such a D-brane configuration
is a ``condensate'' with long-range correlations, blocking up the 
constituents into blocks of different sizes should not change the essential
physics. In what follows, we stick to $2\times 2$ blocks.

\subsection{The commutative limit : matrix renormalization ?}

Consider now an arbitrary function of the noncommutative plane coordinates
(\ref{newnoncoords}), (\ref{newnoncoordsn}) that has a matrix Taylor 
expansion
\be\label{matrixfn}
f({\tilde X}^1,{\tilde X}^2) = \sum_{a_m} f_{a_1a_2\ldots a_m} \
{\tilde X}^{a_1} {\tilde X}^{a_2} \ldots {\tilde X}^{a_m}, \qquad \qquad
a_i=1,2
\ee
There are operator ordering ambiguities in such an expansion for any
nonzero $\lambda\sim \al^{n/2}={1\over 2^{n/2}}$, $\al$ being the averaging 
constant, since ${\tilde X}^1$ and ${\tilde X}^2$ do not commute. With
successive blocking up iterations, $\lambda$ shrinks. Thus the above 
expansion collapses as $\lambda\ra 0$ to
\bea\label{fncommlimit}
f &=& \sum_{a_m} f_{a_1a_2\ldots a_m} \
({x^{a_1}{\bf 1}+D^{a_1}})\ ({x^{a_2}{\bf 1}+D^{a_2}})\ \ldots 
\ ({x^{a_m}{\bf 1}+D^{a_m}})
\\
&\ra& \sum_{m,n} F_{mn} (x^1{\bf 1}+D^1)^m (x^2{\bf 1}+D^2)^n = 
f\biggl((x^1{\bf 1}+D^1),(x^2{\bf 1}+D^2)\biggr) \nonumber\\
&=& f(X^1,X^2)|_{\rm commutative}
\eea
since the diagonal matrices $(x^i{\bf 1}+D^i)$ commute with each 
other\footnote{We have absorbed the $\theta^{12}$ dependence into the 
definition of the diagonal matrices $D^i$ while leaving it as it is in
the off-diagonal ones.}. Thus any well-behaved function asymptotes to 
the commutative restriction of the function in the limit of blocking up
ad infinitum.\\
Further the distance between points 
$P^k=(x^{k 1}{\bf 1}+D^1,x^{k 2}{\bf 1}+D^2)$ is
\be
d_{P^a,P^b} = |(x^{a 1}{\bf 1}+D^1 - x^{b 1}{\bf 1}-D^1)^2 + 
(x^{a 2}{\bf 1}+D^2 - x^{b 2}{\bf 1}-D^2)^2|^{1/2} \equiv |\vx^a - \vx^b|,
\ee
\ie\ the $D^i$ do not show up in the distances between points in the
commutative limiting space.\\
Thus it appears that we cannot really tell the difference between 
$(x^1)$, $(x^1{\bf 1})$ and $(x^1{\bf 1}+D^1)$ in the commutative limit,
insofar as functions with a matrix-Taylor expansion are concerned since
they all collapse down to expressions involving purely commuting 
quantities. In other words, if we replace $(x^i{\bf 1}+D^i)$ by 
$(x^i{\bf 1})$, the functional expression (\ref{matrixfn}), 
(\ref{fncommlimit}) does not appear to change in the limit of an infinite
number of iterations, \ie\ from the point of view of the limiting 
commuting matrices. It is thus interesting to ask if the $D^i$ are 
observable at all and if one can formally define new commuting 
coordinates by shifting away the diagonal matrices $D^i$.
This is essentially equivalent to formally subtracting off the diagonal
matrices $D^i$ and thus the counterterms $\Delta \theta^{12}$ as well. 
This is a $\theta$-dependent ``renormalization'' of the commutative
matrix coordinates. Recall that the $D^i$ are diagonal matrices with 
diverging entries in the limit of infinite blocking up iterations. If this
shift by an infinite diagonal matrix is not observable, this redefinition
in terms of new commuting coordinates $(x^i{\bf 1})$ would be legal.
The diagonal matrices $D^i$ generated reflect the fact that we have lost
information in averaging. The above matrix ``renormalization'' then means
that we have added new counterterms that simulate the effects of 
averaging over short distance open string modes to cancel the generated
counterterms, in much the same way as counterterms in field theory 
generated on changes of an ultraviolet cutoff reflect short distance 
physics that has been integrated out. 
\\
The key physical question is to understand what physical observables are
represented by these classes of matrix functions where shifting away the
$D^i$ is legitimate after blocking up. In other words, when shifting away
the $D^i$ is observable. For instance, it is not clear if shifting by the
$D^i$ is legal at each iteration order. At any finite iteration order, we
have not averaged over all the constituent D0-branes at any point of the 
2-brane condensate so there is a physical difference between the 
$(x^i{\bf 1}+D^i)$ since for example, 
\be\label{diffDi}
(x^i{\bf 1}+D^i)_{nn}-(x^i{\bf 1}+D^i)_{n'n'}=
(D^i)_{nn}-(D^i)_{n'n'} \neq 0
\ee
corresponds, in some sense, to the difference in locations of the $n$ and 
$n'$ 0-branes at the point $(x^1{\bf 1},x^2{\bf 1})$. Thus differences in 
the $D^i$ elements are visible at a finite iteration order. This is 
reminiscent of divergent sums of zero point energies in free field theory 
-- they are unobservable in empty space but do have observable consequences 
in the presence of boundaries as in the Casimir effect for instance. It 
is interesting to look for Casimir-effect-like thought experiments where 
differences (\ref{diffDi}) in the $D^i$s might be observable.
\\
To summarize, we have found that the off-diagonal parts of the algebra
after blocking up are self-similar with reduced noncommutativity
parameter if we subtract off certain ``counterterm'' matrices which
cancel the $D^i$ matrices. \cite{knrp0309} describes a block-spin-like
transformation on some classes of orbifold quiver gauge theories obtained 
by sequentially Higgsing the gauge symmetry in them, thereby performing a
sequence of partial blowups of the orbifolds. It is interesting to
note that subtracting a similar set of diagonal matrices after 
coarse-graining the ``upstairs'' matrices of the ``image'' branes in the 
$C^2/Z_N$ quiver reproduces the worldvolume Higgsing that partially 
blow-up the geometry into a $C^2/Z_{N/2}$ quiver \cite{knrp0309}. More 
generally in the other $C^3/Z_N$ quivers considered there, the 
``counterterm'' matrices required to be subtracted correspond to fields 
that become massive under the Higgsing. Thus perhaps such ``counterterm''
matrices are expected in general to make sense of this matrix 
coarse-graining.
\\
Let us now consider multiplication of two (noncommuting) operators $\cO_f$
and $\cO_g$ on the noncommutative plane corresponding to (commuting) 
functions $f(x)$ and $g(x)$ in the commutative limit. The operators can 
be defined as
\be
\cO_f(X) = {1\over (2\pi)^2} \int d^2k \ {\tilde f}(k) e^{-ik \cdot X}
\ee
where the $X$ are noncommuting coordinates satisfying (\ref{ncplanealg})
and $\ {\tilde f}(k)=\int d^2x\ f(x) e^{ik \cdot x}\ $ is the Fourier 
transform of the (commutative) function $f(x)$.
Then we have $\cO_f \cdot \cO_g = \cO_{f*g}$, where $*$ is the 
usual associative Moyal product that gives
\bea
f*g(x) &\equiv& e^{{i\over 2}\theta^{\mu \nu} \del_{\mu}^y \del_{\nu}^z} 
f(y) g(z)
\bigg|_{y=z=x} = f(x) g(x) + {i\over 4}\theta^{\mu \nu} (\del_{\mu}f 
\del_{\nu}g - \del_{\nu}f \del_{\mu}g) + {\cal O}(\theta^2) \nonumber\\
&=& \int d^2k\ d^2k'\ {\tilde f}(k) {\tilde g}(k'-k)\ e^{-ik' \cdot x}
\ e^{-{i\over 2}\theta^{\mu \nu} k_{\mu} k'_{\nu}}
\eea
Under blocking up, we generate new noncommuting coordinates 
(\ref{newnoncoords}), (\ref{newnoncoordsn}) so that the new operators 
become 
\be
\cO_f = \int d^2k \ {\tilde f}(k) e^{-ik \cdot {\tilde X}}
= \int d^2k \ {\tilde f}(k) e^{-ik \cdot (x{\bf 1}+D+\lambda X)}
\ee
where we recall that $\lambda\sim{1\over 2^{n/2}}$. In the limit of an
infinite number of iterations, $\lambda \ra 0$ and we recover $\cO \ra f$,
the commutative function. Now the operator product becomes 
\bea\label{opmultafter}
\cO_f \cdot \cO_g &=& \int d^2k\ d^2k' \ {\tilde f}(k) {\tilde g}(k')\
e^{-ik \cdot (x{\bf 1}+D+\lambda X)} e^{-ik' \cdot (x{\bf 1}+D+\lambda X)}
\nonumber\\
&\sim& \int d^2k\ d^2k' \ {\tilde f}(k) {\tilde g}(k')\
e^{-i(k+k') \cdot (x{\bf 1}+\lambda X)} \cdot 
e^{-{i\over 2}\lambda^2 \theta^{\mu \nu} k_{\mu} k'_{\nu}}
\eea
where $\sim$ means equality modulo shifting away the $D^i$. Then 
(\ref{opmultafter}) is operator multiplication with a scaled down 
$*$-product, \ie\ reduced noncommutativity. In the limit, 
$\theta^{12}\ra 0$ and we recover commutative multiplication of functions.
\\
Thus, modulo shifting away the $D^i$, the limit of the blocking up 
tranformation is commuting matrices -- 
in some sense, blocking up seems to uncondense the underlying branes,
undoing the noncommutativity, thus moving towards the classical 
commutative limit in a nontrivial way. Note however that it was imperative
that the normalization constant $\alpha$ was fixed relative to the 
commutative limit, specifically demanding that the identity matrix be 
preserved -- an arbitrary $\alpha$ can rescale $\theta^{12}$ to equally
well increase or decrease it.\\
The ``points'' (\ref{ncpoints}) satisfy the algebra (\ref{ncplanealg}) 
and can thus only be defined upto a certain accuracy
\be
\Delta X^1 \ \Delta X^2 \geq \theta^{12},
\ee
where the uncertainty in a given state $|\psi\rA=\sum_n c_n |n\rA$
is defined as usual as the variance 
$(\Delta X^i)^2=\lA \psi|{X^i}^2|\psi\rA - (\lA \psi|X^i|\psi\rA)^2$.
After $n$ blocking up iterations (with $2\times 2$ blocks), 
$\theta^{12} \ra {\theta^{12}\over 2^n}$.
Thus the uncertainty shrinks under repeated block transformations, if 
we hold the distances between the commutative coordinates fixed.
In the limit with commuting matrices, the uncertainty vanishes and we
have ``points'' effectively defined with arbitrary resolution. 
It would be very interesting to study if the change in this resolution
can be understood along the lines of Wilsonian effective actions -- for
example, in terms of lowering an ultraviolet cutoff in momenta and 
lowering the sensitivity of a field theory to high energy processes.\\
It is noteworthy that since several noncommuting matrices give rise to 
the same diagonal commuting matrices, the set of commuting matrices is 
a coarser description -- blocking up is irreversible and we have lost 
information in blocking up.

\subsection{Fluctuations : noncommutative field theory under blocking up}

We have been working with $\al'\ra 0$ to decouple the higher massive
string modes, enabling the restriction to field theory. We keep $g_s$
finite so as to retain interactions in the field theory \cite{sw9908}, 
\cite{Seiberg0008}. Consider a single infinite flat D2-brane made up 
of an infinite number of D0-branes as a noncommutative plane 
(\ref{ncplanealg}). Small fluctuations about the background 
(\ref{ncplanealg}) are obtained by expanding the leading terms in the 
Born-Infeld action in terms of background independent variables 
$Y^i=X^i+\theta^{ij}A_j$, thereby defining noncommutative gauge field 
strengths and other quantities \cite{sw9908}, \cite{Seiberg0008}. This 
can be rewritten, using $*$-products, as an abelian gauge theory action 
with an infinite series of higher derivative terms arising from the 
$*$-product. Recall that loop corrections to the effective action yield 
new IR poles from the nonplanar contributions to the effective action 
\cite{MRS9912}. These IR poles a priori wreck any Wilsonian interpretation 
of noncommutative field theories since integrating out short distance 
fluctuations gives rise to long distance effects (UV/IR mixing
\cite{MRS9912}, \cite{susskind0002}).\\
Let us assume that the matrices $D^i$ have been formally 
dropped\footnote{It is conceivable that this shift by a diagonal matrix
is in fact equivalent to a noncommutative gauge transformation involving
spatial translations \cite{grossakiitz0008}. To understand this shift by 
the $D^i$ at each iteration 
order, we need to systematically segregate background from fluctuations 
at each iteration order, incorporating possible translational gauge 
redundancies.}. Then under blocking up, 
(\ref{opmultafter}) gives precisely a scaled down $*$-product -- the 
terms involving $D^i$ have been formally dropped. Since the $*$-product
scales down exponentially under blocking up as $\theta^{12} \ra 
{1\over 2^n}\theta^{12}$, the scale of noncommutativity in each of the
higher derivative terms shrinks down, although there still is an infinite
series of these terms. In the limit of an infinite number of blocking up
iterations, we recover an abelian commutative gauge theory. Note that
since $\ \theta\sim{1\over B}\ra 0,\ $ this is an abelian commutative 
gauge theory with $B^{12}\ra \infty$, \ie\ an infinite background magnetic
field. Thus this is not the same commutative theory as the one with no
background gauge fields --- once we turn on a magnetic field and take the
decoupling limit $\al'\ra 0$ to restrict to noncommutative field theory,
it does not appear possible to analytically take the limit $B^{12}\ra 0$
and recover a commutative field theory, essentially due to the order of 
limits taken.
\\
Now in this $\theta^{12}\ra 0$ commutative limit, the IR poles are absent.
Thus it appears that under blocking up, we have somehow integrated out 
the IR poles. Note however that even in the commutative limit, there exist
quantum fluctuations yielding loop corrections to the abelian commutative
gauge theory. Thus it appears that blocking up has somehow integrated out 
precisely the noncommutative fluctuations about the noncommutative plane
background, in other words the nonplanar contributions to the effective
action. The IR poles arose from the nonplanar contributions alone so that
this interpretation seems consistent. It would be interesting to 
understand what precisely the physics is here and if this interpretation
of integrating out the IR poles is correct. For example, formally
dropping the counterterm contributions may have thrown out precisely
the pieces required to bring back in the IR poles, akin to an ``anomaly''
under the process of blocking up branes. If such an ``anomaly'' exists, 
our interpretation above is simply wrong. Clearly a more careful analysis
is required here.
\\
It is not clear if a prescription precisely analogous to a Wilsonian
one of integrating out can be formulated here, since the infinite
series of higher derivative terms after each iteration still yield
nontrivial nonplanar contributions to processes thus retaining UV/IR
mixing at any finite iteration order. However, given that the blocking
process shrinks noncommutativity, it appears that blocking up
generates a flow between noncommutative field theories with different
$\theta^{ij}$. Thus, perhaps the analogs of Callan-Symanzik equations
do exist for the way couplings and parameters in a noncommutative
field theory transform under blocking up. It would be very interesting
to understand these issues better.

\subsection{On the role of gauge invariance}

Firstly it is imperative to note that (\ref{NCsoln}) is a particular
solution to the algebra (\ref{ncplanealg}). It is a priori not obvious 
that blocking up will give any self-similar structure, if this solution
is modified.

Since $X^1,X^2$ do not commute, we could not have simultaneously
diagonalized these matrices by any gauge transformation. 
However it is important to note that the basis states can in principle be
ordered differently -- there exist $U(\infty)$ gauge transformations that
can be used to map new basis states by simply rotating the basis of energy
eigenstates $|n\rA$. This is simply a change of basis and therefore does
not change the physics.
\\
For simplicity, consider the new basis obtained by interchanging two 
states alone while leaving the other states unchanged
\be\label{newbasisorder}
|0\rA = (1,0,0,\ldots)^T  \ \leftrightarrow \ |2\rA=(0,0,1,\ldots)^T,
\ \qquad |n'\rA\equiv|n\rA, \ \ n\neq 0,2
\ee
This is implemented by the $U(\infty)$ gauge transformation
\bea\label{gaugetransfU}
U=
\left( \bA{cccccc} 0 & 0 & 1 & 0 & & \ldots \\ 
0 & 1 & 0 & 0 & & \ldots \\ 
1 & 0 & 0 & 0 & & \ldots \\
0 & 0 & 0 & 1 & 0 & \ldots \\
& & & 0 & 1 &  \\
& & & . & &  \\
& & & . & &  \\
\eA \right)
\eea
Then, for example, the new $X^1$ matrix becomes
\bea\label{newbasisX}
{X^1}' = U X^1 U^{\dag}
= x^1{\bf 1} + \sqrt{\theta^{12}\over 2}
\left( \bA{ccccccc}
0 & \sqrt{2} & 0 & \sqrt{3} & & & \ldots \\ 
\sqrt{2} & 0 & 1 & 0 & & & \ldots \\ 
0 & 1 & 0 & 0 & & & \ldots \\
\sqrt{3} & 0 & 0 & 0 & \sqrt{4} & & \ldots \\
& & & \sqrt{4} & 0 & \sqrt{5} & \\
& & & & \sqrt{5} & 0 & \\
& & & . & & & \\
& & & . & & & \\
\eA \right)
\eea
This shows that a gauge transformation for basis change has transformed 
the $X^1$ matrix from having nonzero entries only in the one-off-diagonal
slot to one with far-off-diagonal nonzero matrix elements. Blocking up 
(\ref{newbasisX}) does not appear to give any self-similar or otherwise
recognizable structure, although it does move towards a near-off-diagonal
form and shrink the off-diagonal modes over many iterations.

Reversing the logic above, it is plausible to expect that, in general, 
basis-changing gauge transformations will transform a matrix representation
such as (\ref{newbasisX}) with far-off-diagonal nonzero matrix elements
into one with as near-off-diagonal a form as possible. Such gauge
transformations could then be used to transform a general matrix
representation of a noncommutative algebra into one which is as 
near-off-diagonal as possible. It is far from obvious however that such
gauge transformations exist for a general noncommutative algebra. Further
it is very possible that there exist other kinds of gauge transformations
that might make it hard to sensibly apply such block-brane techniques.
\\
Note that the order (\ref{basisorder}) in which we have chosen to write 
the basis has automatically ordered the states so that nearest neighbour 
elements increase in numerical value so that the block-averaged elements
also have their values ordered -- this has ensured that averaging is 
physically sensible and indeed seems to correspond in some sense to 
averaging over open string modes stretched between nearest neighbour 
D-branes in physical space. Overall, it appears that blocking up is
sensitive to the gauge choice in general. However this is perhaps not 
terribly outlandish -- indeed, consider the following analogy with a 
1D Ising-like spin-chain with Hamiltonian
\be
H = J \sum_i s_i s_{i+1} = J\ {\vec s}^T\cdot M\cdot {\vec s}
\ee
where $J$ is the spin-spin coupling. We have written the Hamiltonian in
terms of a ``spin vector'' ${\vec s}=(\ldots, s_1, s_2, s_3, \ldots)$,
$M$ being the matrix of spin-interactions. With nearest neighbour 
couplings alone, M is of the form of the $X^i$ matrices, but with equal 
one-off-the-diagonal nonzero entries. Consider now a basis changing 
transformation on the spin vector, ${\vec s}\ra {\vec s}'=U{\vec s}$. 
This does not change the physical system -- just what we chose to call 
spin labels $s_i$ appearing in the Hamiltonian. For concreteness, 
interchanging, say, $s_1\leftrightarrow s_3$ is given by a gauge 
transformation matrix $U$ of the same form as (\ref{gaugetransfU}) 
(insofar as $s_1,s_3$ are concerned, the rest being unchanged). But this 
now transforms the Hamiltonian in the old spin basis to one with 
third-nearest neighbour interactions involving the interchanged spins 
in this new basis
\be
H =  J (\ldots + s_0 s_1 + s_1 s_2 + s_2 s_3 + s_3 s_4 + \ldots)
\ra J (\ldots + s_0' s_3' + s_3' s_2' + s_2' s_1' + s_1' s_4' + \ldots)
\ee
The point here is that an implicit gauge choice seems to be made in 
general in studies of such condensed matter systems, that, \eg, spins in
physical space are identified with spin labels in the Hamiltonian. 
This is a gauge choice that facilitates treatments of blocking up 
nearest-neighbour spins in physical space by allowing us to block up 
nearest-neighbour spin labels in the Hamiltonian. It thus appears that
a similar gauge-fixing is at play in our story with D-branes, suggesting
that there exists a ``physical'' gauge, where D-brane locations in 
physical space are identified with D-brane locations in the elements of
their matrix representations. It would be interesting to understand what 
the corresponding gauge-invariant way of blocking up D-branes and 
matrices is.

\section{Coarse-graining other nonabelian geometries}

\subsection{Membranes : the fuzzy torus and the fuzzy sphere}

Consider a D2-brane described as a fuzzy torus $T^2_q$ given by the 
matrix algebra
\be
U V = q V U, \qquad \qquad q=e^{2\pi i/N}
\ee
with a representation in terms of the $N\times N$ clock-shift matrices 
\bea
U_q=\left( \bA{ccccc} 1 & 0 & & \ldots & \\ 
0 & q & 0 & \ldots & \\ 
& 0 & q^2 & \ldots & \\ 
& & . & & \\ 
& & . & & \\ 
& & & 0 & q^{N-1} \\ 
\eA \right),
\qquad \qquad
V_q=\left( \bA{ccccc} 0 & 1 & & \ldots & \\ 
& 0 & 1 & \ldots & \\ 
& & 0 & 1 \ldots & \\ 
& & . & & \\ 
& & . & & \\ 
1 & & & & 0 \\ 
\eA \right)
\eea
Under $(2\times 2)$ blocking up with a uniform averaging constant $\al$, 
these matrices transform into ${N\over 2}\times {N\over 2}$ matrices 
\be
N\ra {N\over 2}, \qquad q \ra q^2=e^{2\pi i/(N/2)} \qquad
U_q \ra \al (1+q) U_{q^2}, \qquad V_q \ra \al ({\bf 1}+V_{q^2})
\ee
which clearly satisfy a similar fuzzy torus algebra of lower rank, as 
before upto expressions involving diagonal ``counterterm'' matrices. 
Thus $T^2_q$ coarse-grains to $T^2_{q^2}$.

Now consider a fuzzy sphere formed from $N$ D0-branes stabilized by
the presence of a 4-form field strength $f$ \cite{myers99}. This can 
be described as a spherical membrane \cite{kabattaylor} using the 
familiar angular momentum $SU(2)$ Lie algebra
\be\label{XJsphere}
X_i=f J_i, \qquad \qquad
[J_i, J_j] = i \epsilon_{ijk} J_k, \qquad i, j, k = 1,2,3
\ee
This has finite dimensional representations with the $X^i$ satisfying
\be
X_1^2+X_2^2+X_3^2=f^2 J^2=R^2{\bf 1}
\ee
so that the sphere has radius $R=fN$, upto $O(1/N)$ corrections.\\
Defining the usual raising and lowering operators from $J_1,J_2$ as
$J_{\pm} = {1\over \sqrt{2}} (J_1 \pm i J_2)$, the algebra can be put in
the form $\ [J_3, J_{\pm}] = \pm J_{\pm}, \ [J_+, J_-] = J_3\ $
so that the representation matrices of $J_1, J_2$ will only have nonzero 
next-to-diagonal matrix entries while $J_3$ is diagonal. Thus we can expect
that blocking up will be somewhat similar to that for the noncommutative
plane.
Specifically, states of the algebra can be labelled by eigenstates of 
$J^2$ and $J_3$ as $\ J^2 |j,m> = j(j+1) |j,m>, \ J_3 |j,m> = m |j,m>$.
In this basis, we can write the matrix elements of $J_1, J_2$ as
\bea
<m| J_1 |m-1> &=& \ \ <m-1| J_1 |m> = {1\over 2} \sqrt{(j+m)(j-m+1)}, 
\nonumber\\
<m| J_2 |m-1> &=& -<m-1| J_2 |m> = -{1\over 2}i \sqrt{(j+m)(j-m+1)}, 
\eea
Let us focus for convenience on the $2j+1=2N$ representation. $m$ takes
values from $-(N-{1\over 2})$ to $(N-{1\over 2})$. Then $J_1$ has the 
representation matrix 
\bea
\left( \bA{cccccccc} 0 & {\sqrt{2N-1}\over 2} & 0 & & & & & \ldots \\ 
{\sqrt{2N-1}\over 2} & 0 & {\sqrt{2(2N-2)}\over 2} & 0 & & & & \ldots \\ 
0 & {\sqrt{2(2N-1)}\over 2} & 0 & {\sqrt{3(2N-3)}\over 2} & 0 & & & \ldots \\
& 0 & {\sqrt{3(2N-3)}\over 2} & 0 & {\sqrt{4(2N-4)}\over 2} & 0 & & \ldots \\
& & 0 & {\sqrt{4(2N-4)}\over 2} & 0 & {\sqrt{5(2N-5)}\over 2} & 0 & \\
& & & & {\sqrt{5(2N-5)}\over 2} & 0 & {\sqrt{6(2N-6)}\over 2} &  \\
& & & & . & & &  \\
& & & & . & & &  \\
\eA \right)
\eea
while that for $J_2$ can be written in a similar fashion and $J_3$ has
the representation matrix
\be
J_3 = {\rm diag}\ [-(2N-1)/2,-(2N-3)/2,\ldots,(2N-3)/2,(2N-1)/2]
\ee
Then it is easy to see that the matrices coarse-grain under ($2\times 2$)
blocking as
\be
J_1^{(N)} \ra 2\al J_1^{(N/2)}, \qquad J_2^{(N)} \ra 2\al J_2^{(N/2)}, 
\qquad J_3^{(N)} \ra 2\al_3 J_3^{(N/2)}
\ee
We see that demanding a self-similar structure (upto diagonal
``counterterm'' matrices) in this case requires the averaging constants 
to be non-uniform $\al_3 = {\al\over 2}$, somewhat similar to the
coarse-graining for the quiver examples in \cite{knrp0309}. Note that
if we hold the sphere radius $R=fN$ fixed, then as $N\ra {N\over 2}$ 
the 4-form field strength rescales as $f\ra 2f$ under this 
coarse-graining (recall that reduced noncommutativity in the 
noncommutative plane implied that the B-field rescaled). In general, 
since closed string backgrounds appear as parameters in the effective 
D-brane configuration, we expect that the closed string backgrounds 
also rescale under such a coarse-graining of the D-brane configuration.
It would be interesting to understand the structure of the flows that 
are obtained in general under this coarse-graining.
\\
We have assumed here that $N$ is even, which is reasonable for large $N$.
We expect that membranes with other topologies will exhibit similar
structures upto $O(1/N)$ corrections. It is also easy to see that the
block size does not matter in these examples as well, just as in the 
noncommutative plane.
Since the longitudinal momentum in Matrix theory is proportional to $N$,
shrinking $N$ would appear to suggest that we are in some sense carrying 
out a large $N$ light-front renormalization. It would be interesting to 
understand this in detail.

\subsection{Four-branes}

D4-branes with nonzero membrane charge can be constructed as in 
\cite{banksseibergshenker} in terms of multiple noncommutative 2-planes,
as \eg
\be
[X^1,X^2]=i\theta^{12}{\bf 1}, \qquad
[X^3,X^4]=i\theta^{34}{\bf 1}
\ee
with the other commutators vanishing. Then in addition to nonzero 
2-brane charges along the $12$ and $34$ planes, we also have nonzero 
4-brane charge
\be
{\rm Tr}\ \epsilon_{abcd} X^a X^b X^c X^d \sim \theta^{12} \theta^{34}
\ee
It is clear then that taking a representation in terms of a 
4-dimensional phase space spanned by two sets of creation-annihilation 
operators $a_1,a_1^{\dag},a_2,a_2^{\dag}$, the methods of coarse-graining
apply in the two independent 2-planes. It would be interesting to study
other brane configurations along these lines.

\section{Blocking up in general}

More generally, let us ask what happens to a general matrix configuration
representing a given D-brane background, with far-off-diagonal matrix
entries turned on. Gauge transformations can of course modify a matrix
element from being far-off diagonal to near-off-diagonal. For instance,
there exist gauge transformations that change the ordering of the energy 
eigenstates in the noncommutative plane. As we have seen, the 
representation matrices then contain far-off-diagonal elements. We 
therefore assume that there exist, as in the noncommutative plane, gauge
transformations that transform a generic representation of 
(\ref{branebgnd}) into one that has as near-off-diagonal a form as 
possible. It is far from obvious what physics underlies, in common, those 
D-brane configurations which can be put in as near-off-diagonal a form
as possible using such gauge transformations. In what follows, we simply 
make this nontrivial assumption without justification and work out the
consequences.

Assuming that the matrix can be put in as near-off-diagonal a form as 
possible is equivalent to assuming that matrix entries beyond a 
certain off-diagonal slot vanish uniformly along the matrix, so that the
matrix ``truncates'' a finite (albeit perhaps large) distance from the
diagonal (see figure~\ref{fig2}). Such a finite matrix truncation is 
analogous to a finite (albeit perhaps large) correlation length in 
critical phenomena. Blocking up the constituent spin degrees of freedom 
then yields a system with a small correlation length which then has a 
more tractable Hamiltonian. This translates by analogy to saying that 
matrix truncation is expected to move inwards, towards smaller ``matrix
correlation lengths'' under blocking up. In what follows, we shall see
that this is indeed the case.
\\
It is noteworthy that this does not however give insight into the 
case of strictly infinite correlation lengths. Naively since matrices
that do not have a finite truncation after any gauge transformation 
continue to retain an infinite number of off-diagonal matrix entries,
strictly infinite matrix correlation lengths appear to be best 
interpreted as nontrivial fixed points of matrix renormalization.

\subsection{Some pictorial observations}

\begin{figure}
\bc
\epsfig{file=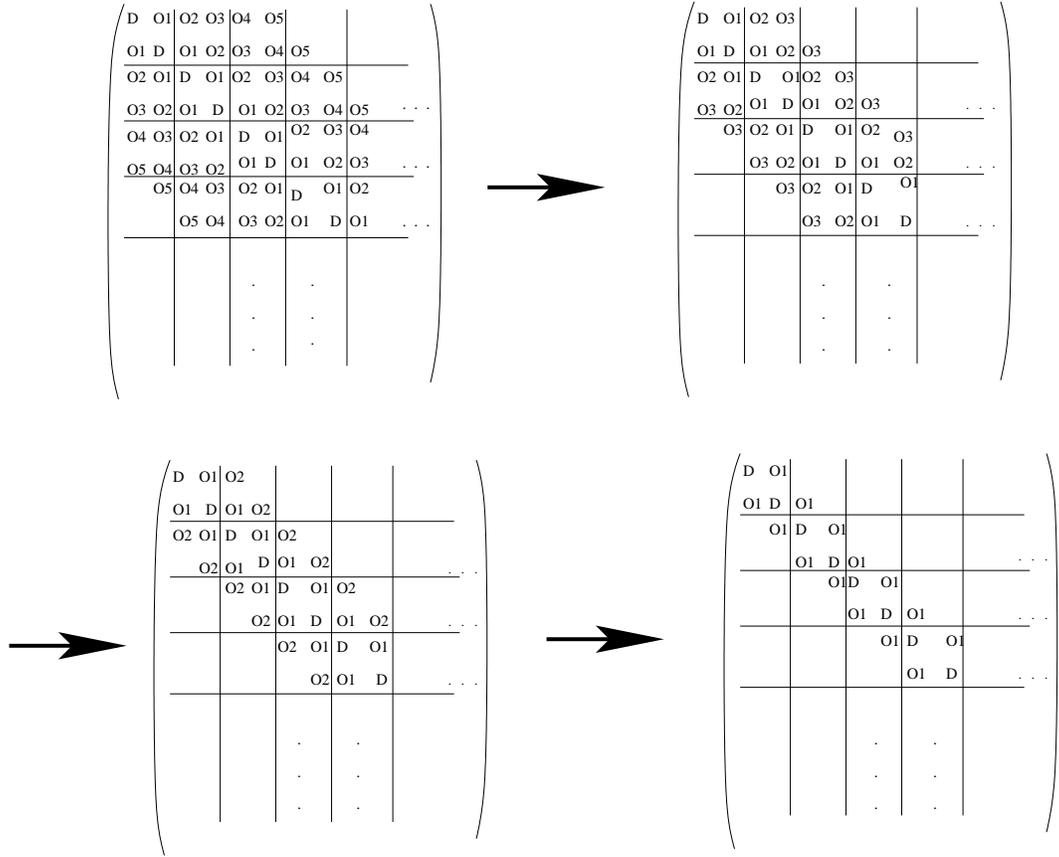, width=14cm}
\caption{Blocking up an $\{ O5 \}$ original matrix flows successively
towards a near-off-diagonal blocked-up matrix, specifically through
$\{ O3 \}$, $\{ O2 \}$ and finally $\{ O1 \}$, which remains 
$\{ O1 \}$.}
\label{fig2}
\ec
\end{figure}

We recall that the scalar matrices are hermitian matrices in the adjoint
of some gauge group (which we take for concreteness to be $U(\infty)$).
Thus the matrices are (pictorially) symmetric about the diagonal. 
Consider, as in figure~\ref{fig2}, a general matrix under blocking up.
$\{ {\rm D} \}$ schematically represents a possibly nonzero diagonal 
matrix entry while $\{ O \nu \}$ schematically represents a possibly 
nonzero entry in the slot that is $\nu$ off the diagonal. As described 
above, we assume that after appropriate gauge transformations, the 
matrix ``truncates'' a finite distance from the diagonal. For purposes
of illustration, we have assumed in figure~\ref{fig2} that matrix
entries beyond $\{ O5 \}$ vanish. It is clear that starting with 
$\{ O5 \}$ as the furthest nonzero entry, blocking up has brought the 
furthest nonzero entry inwards, towards the diagonal -- $\{ O3 \}$ is 
the furthest nonzero off-diagonal entry after one iteration. Iterating
once more gives the furthest off-diagonal entry as $\{ O2 \}$. Yet 
another iteration finally gives a matrix whose furthest off-diagonal 
entry is $\{ O1 \}$. We have seen in the analysis of the noncommutative
plane example that such an $\{ O1 \}$ is a ``fixed point'' of the 
blocking up transformation (at this pictorial level).

Indeed consider $k\times k$ blocks with the furthest nonzero off-diagonal
entry being in the $n$-th column. Then after blocking up, the furthest
nonzero off-diagonal entry lives in the $n'=[{n+k-1\over k}]$-th block,
where we understand this expression to be the nearest greater integer. 
Then the fixed points of this iterative equation are given by $n' = n$ 
which reduces to
\be
n-1 = \biggl[ {n-1 \over k} \biggr] \ \ra_{k \ {\rm large}} \ 1
\ee
are $n-1 = 1$, since by taking large enough $k$, the right hand side is a
fraction which truncates to unity. In other words, $n=2$, \ie\ $\{ O1 \}$,
is a fixed point. $n=1$, \ie\ $\theta=0$, is of course a fixed point.

It thus appears pictorially that blocking up tends to average over
far-off-diagonal modes and bring them in towards the diagonal and stop 
at order $\{ O1 \}$. Let us refer to $\{ O \nu \}$ as the degree of 
noncommutativity exhibited by the background matrices $X^i$, after 
appropriate gauge fixing. Then a matrix with far-off-diagonal entries
turned on can be referred to as strongly noncommutative while a 
background of degree, say $\{ O1 \}$, is weakly noncommutative. Then 
from above, we see that under blocking up, backgrounds seem to ``flow''
towards weak noncommutativity. Indeed, in general, besides purely 
diagonal matrices, \ie\ of type $\{ O0 \}$, we see pictorially that 
$\{ O1 \}$ and $\{ O\infty \}$ matrices are also possible ``fixed 
points'' of the blocking up transformation. 

It is interesting that the set of $\{ O1 \}$ structures is rather
large and allows ample opportunity for interesting physics. Indeed, one
expects that all Lie algebras can be subsumed into this set by writing
the Lie algebra as a Cartan subalgebra with additional raising-lowering
generators (these would only connect nearest neighbour states of the 
representation\footnote{There is in fact some independent evidence for 
this \cite{vvs0010}.}, hence $\{ O1 \}$). In general, one expects that 
nonlinear $\theta^{ij}$ would give rise to strongly noncommutative 
geometries, \ie\ $\{ O\nu \}, \ \nu>1$ structures : $\theta^{ij}$ with 
linear dependence on the $X^i$s are in general expected to give rise to 
$\{ O1 \}$ algebraic structures such as Lie algebras.

In the above pictorial analysis, we have not studied the detailed values
of the matrix elements, just how matrix truncation scales overall. 
In general, the matrix entries themselves are expected to get 
``renormalized'' and may or may not reflect any detailed self-similarity
under blocking up -- the detailed evolution of the matrix under blocking
up depends on the specific algebra in question and at a deeper level,
the underlying geometric structure. In the next subsection, we study in
more detail some algebraic aspects of blocking up. We describe some
heuristic scaling relations for the matrix elements and study the 
counterterms generated under this coarse-graining.

It is amusing to note in passing that the observations of this pictorial
subsection are, on the face of it, applicable as approximation schemes
to matrix-like structures anywhere.

\subsection{Some algebraic properties}
\subsubsection{Scaling relations for shrinking ``off-diagonality''}

Consider the blocked matrix elements (\ref{defnblock}) for 
$2\times 2$ blocks 
\be
{\tilde X}^i_{a,b} = \alpha (X^i_{2a-1,2b-1} + X^i_{2a,2b-1} 
+ X^i_{2a-1,2b} + X^i_{2a,2b}) ,
\qquad \qquad a,b=1 \ldots {N\over 2} \nonumber
\ee
We are assuming that $N$ is essentially infinite here, as mentioned 
previously. From the previous subsection, we have seen that successive 
block transformations flow towards $\{ O1 \}$ matrices. Let us therefore 
restrict attention to the special case where $\{ O1 \}$ is the furthest 
nonzero off-diagonal element, \ie\ at most $X^i_{nn}, X^i_{n,n\pm 1} 
\neq 0$. Then at most ${\tilde X}^i_{nn}, {\tilde X}^i_{n,n\pm 1} \neq 0$ 
under blocking up. Indeed we have as nonzero elements,
\bea
{\tilde X}^i_{a,a} &=& \alpha (X^i_{2a-1,2a-1} + X^i_{2a,2a-1} 
+ X^i_{2a-1,2a} + X^i_{2a,2a}), \nonumber\\
{\tilde X}^i_{a,a+1} &=& \alpha (X^i_{2a-1,2a+1} + X^i_{2a-1,2a+2} 
+ X^i_{2a,2a+1} + X^i_{2a,2a+2}) = \alpha X^i_{2a,2a+1}, \nonumber\\
{\tilde X}^i_{a,a-1} &=& \alpha (X^i_{2a-1,2a-3} + X^i_{2a,2a-3} 
+ X^i_{2a-1,2a-2} + X^i_{2a,2a-2}) = \alpha X^i_{2a-1,2a-2},
\eea
and their hermitian conjugates --- the remaining vanish. For a general 
geometry, there is no physical reason to demand that the identity matrix
be preserved, so that in general $\al\neq{1\over 2}$ -- we will therefore
keep $\al$ as a floating variable here. 
Recall that we demand $\alpha<1$, in accordance with what we intuitively 
expect of averaging. It is interesting to look for the conditions under
which the blocked off-diagonal modes are smaller than the corresponding
modes of the original matrix
\be
\biggl| {{\tilde X}^i_{a,a+1}\over X^i_{a,a+1}} \biggr| =
\biggl| {\alpha X^i_{2a,2a+1}\over X^i_{a,a+1}} \biggr| < 1
\ee
After $n$ iterations, the analog of this equation gives
\be\label{commlimitcondn1}
| {\tilde X^{i (n)}}_{a,a+1} | = 
| \alpha^n \ X^{i (0)}_{2^n.a,2^n.a+1}| < | X^{i (0)}_{a,a+1} |
\ee
Thus the condition that blocking up shrinks the off-diagonal modes 
thereby reducing noncommutativity is that there exists some $\alpha$ 
such that this inequality holds for all $a$ labelling the off-diagonal 
modes, after $n$ iterations, for all $n$, in whichever basis
we choose to write the matrices (it is easy to generalize this to 
$k\times k$ blocks). Roughly speaking, this condition says that the 
elements of the original ``seed'' matrix $X^{i (0)}$ must not grow too 
fast down the matrix if blocking up is to move towards a commutative 
limit. For the geometries where there is a physical reason for the 
identity to be preserved under blocking up, we have $\al=1/2$ in the 
expressions here. In general, (\ref{commlimitcondn1}) suggests the 
existence of a ``critical'' value of $\al$ that is required for a given 
noncommutative geometry to move towards the commutative limit under 
blocking up.
\\
In the case of the noncommutative plane, we see that this condition
was satisfied in the ordered energy eigenstate basis
(\ref{basisorder}) that we expressed the matrices in. This suggests
that that basis was, in a sense, a ``physical'' basis amenable to
blocking up.  The condition (\ref{commlimitcondn1}) thus appears to be
a nontrivial requirement on the expectation values of the scalar 
matrices allowed for blocking up to shrink noncommutativity. It is 
important to note that this analysis is restricted in some sense to 
local properties of the geometry. Global issues might well wreck this 
framework since there could be nontrivial topological obstructions to 
recovering smooth limits. We will return to this point in Sec.~6.
\\
Similarly, the conditions under which the blocked-up off-diagonal
modes are smaller than the corresponding diagonal modes (we give 
only the conditions after the first iteration, the later iterations
can be written easily enough) can be written as
\be \label{commlimitcondn2}
\biggl| {{\tilde X}^i_{a,a+1}\over {\tilde X}^i_{a,a}} \biggr| =
\biggl| {X^i_{2a,2a+1}\over (X^i_{2a-1,2a-1} + X^i_{2a,2a} 
+ X^i_{2a-1,2a} + X^i_{2a,2a-1})} \biggr| < 1
\ee
Assuming this condition to be true is self-consistent -- if we assume 
that diagonal elements are dominant relative to the off-diagonal ones, 
then blocking up preserves this assumption. Indeed, consider the heuristic
scaling relation
\be\label{scalingrelation}
X^i_{a,b} \sim X^i_{a,a}\ q^{|b-a|}, \qquad \qquad {\rm for \ some} \ |q|<1
\ee
Then assuming $X^i_{c-1,c-1}\sim X^i_{c,c}$, (\ref{commlimitcondn2}) 
gives
\be
\biggl| {{\tilde X}^i_{a,a+1}\over {\tilde X}^i_{a,a}} \biggr| 
\sim \biggl| {q \over 2(1+q)} \biggr| \equiv |{\tilde q}| < |q|
\ee
giving rise to a similar scaling relation as (\ref{scalingrelation})
\footnote{Note that if the averaging constant $\al={1\over 2}$, we 
have ${\tilde q}\sim \al q$.}. More generally, using (\ref{defnblock}),
the scaling relation (\ref{scalingrelation}) gives for the block-matrix 
elements
\bea
{\tilde X}^i_{a,a} &\sim& \al\ X_{2a,2a}\ (1+q+1+q) \sim 
\al\ 2(1+q)\ X_{2a,2a},
\nonumber\\
{\tilde X}^i_{a,b} &\sim& \al\ X_{2a,2a}\ 
(q^{2(b-a)}+q.q^{2(b-a)}+q^{-1}.q^{2(b-a)}+q^{2(b-a)})
\sim \al\ X_{2a,2a}\ q^{2(b-a)}\ {(1+q)^2\over q} \ \ \qquad
\eea
so that the new scaling relation is
\be
{\tilde X}^i_{a,b} \sim {\tilde X}^i_{a,a}\ {\tilde q}^{|b-a|}
\ee
where 
\be
{\tilde q} \sim q^2\ \biggl({1+q \over 2q}\biggr)^{1\over |b-a|}
\ee
It is interesting to note that $\ {\tilde q} \sim q^2$ for $b\gg a$, 
\ie\ far-off-diagonal modes, which thus scale down fast. On the other
hand, ${\tilde q} \sim q/2$ for, \eg\ $b=a+1$, \ie\ the one-off-diagonal 
modes. Thus the heuristic scaling relation (\ref{scalingrelation}) shows
that under this coarse-graining, the matrices in general become 
``less off-diagonal''. It is noteworthy that ${\tilde q}\sim q$ if 
$q\sim O(1)$ in this scaling relation, thus suggesting the potential 
emergence of a fixed point if the off-diagonal modes are as dominant as 
the near-diagonal ones.

\subsubsection{Matrix invariants as possible c-functions}

It is interesting to ask if there is any quantity that monotonically
decreases under this coarse-graining, thus behaving like a possible
c-function. It is hard to address this question in the context of 
several noncommuting matrices but focussing on commuting matrix 
representations gives some insight.\\
Consider one Hermitian $N\times N$ matrix $M$, which thus can always be 
diagonalized to the form 
$\ M={\rm diag}\ [\lambda_1,\lambda_2,\ldots,\lambda_N]$, 
where the $\lambda_i$ are the $N$ eigenvalues. One can construct the 
$N$ gauge-invariant observables 
\be
{\rm Tr} M^k = \sum_{i=1}^N \lambda_i^k
\ee
Under $(2\times 2)$-block coarse-graining, we have 
$\ M\ra 
{\tilde M}=\al\ {\rm diag}\ [\lambda_1+\lambda_2,\lambda_3+\lambda_4,
\ldots,\lambda_{N-1}+\lambda_N]$. Then the new observables are
\be
{\rm Tr} {\tilde M}^k = \al^k \sum_{i=1}^{N/2} 
(\lambda_{2i-1}+\lambda_{2i})^k
\ee
It is easy to see that the $k=2$ observable can be reorganized and 
compared with its expression before coarse-graining as
\bea
{\rm Tr} {\tilde M}^2 - {\rm Tr} M^2 
&=& \al^2 (\lambda_1^2 + \lambda_2^2 + 2 \lambda_1 \lambda_2) -
(\lambda_1^2 + \lambda_2^2) + \ldots
\nonumber\\
{} &=& -(1-\al^2)\biggl[ \biggl(\lambda_1 - {\al^2\over 1-\al^2}\lambda_2
\biggr)^2 + (1-{\al^4\over (1-\al^2)^2}) \lambda_2^2 \biggr] + \ldots
\eea
This expression for $\al<1$ is nonpositive if $\al^2 \leq {1\over 2}$.
Thus for $\al$ within this critical value, $\ {\rm Tr} M^2$ is 
monotonically non-increasing, akin to a c-function for this 
coarse-graining. It would be interesting to understand generalizations
of this calculation.

\subsubsection{Counterterms}

With an $\{ O1 \}$ structure, keeping only the nonzero terms in 
expanding $\theta^{ij}_{ab}$ gives
\be
\theta^{ij}_{a,b} = 
X^i_{a,a-1} X^j_{a-1,b} + X^i_{a,a} X^j_{a,b} + X^i_{a,a+1} X^j_{a+1,b}
- (i \leftrightarrow j)
\ee
so that $\theta^{ij}_{ab}$ is nonzero only for $b=a,(a\pm1),(a\pm2)$. 
For \eg,
\bea \label{theta0}
\theta^{ij}_{a,a+1} &=& X^i_{a,a} X^j_{a,a+1} + X^i_{a,a+1} X^j_{a+1,a+1}
- (i \leftrightarrow j) \nonumber\\
{} &=& X^i_{a,a+1} (X^j_{a+1,a+1} - X^j_{a,a}) - (i \leftrightarrow j)
\sim \ {\cal O}(\{ O1 \}) [{\rm mass \ term}]
\eea
where ``mass term'' is essentially the difference in vevs (akin to a 
Higgs mass arising from separated D-branes).
There are also corresponding antihermitian conjugate elements
(by definition, $\theta^{ij}_{b,a}=-{\theta^{ij}}^*_{a,b}$).
\\
Let us now calculate the commutator of the ${\tilde X}^i$s using the 
definition (\ref{defnblock})
\bea
\biggl[ {\tilde X}^i, {\tilde X}^j \biggr]_{a,b} 
&=& \al^2 \sum_{c=1}^{N/2} \biggl[ 
\biggl( X^i_{2a-1,2c-1} + X^i_{2a-1,2c} + 
X^i_{2a,2c-1} + X^i_{2a,2c} \biggr) \cdot \qquad \nonumber\\
&& {} \qquad \qquad \biggl( X^j_{2c-1,2b-1} + X^j_{2c-1,2b} + 
X^j_{2c,2b-1} + X^j_{2c,2b} \biggr) - (j \leftrightarrow i) \biggr]
\eea
On expanding the brackets and grouping the terms, one finds after a little
algebra
\be
\biggl[ {\tilde X}^i, {\tilde X}^j \biggr]_{a,b} 
= i \ \alpha \biggl( \theta^{ij}(X) \biggr)^{\rm blocked-up}_{a,b}
+ \Delta \theta^{ij}_{a,b}(X)
\ee
The $\Delta \theta$ are counterterms, as in the noncommutative plane.
It is clear that if these counterterms vanish, the algebra 
(\ref{branebgnd}) of the $X^i$s is self-similar under blocking up. Now 
if $\alpha$ is chosen appropriately, blocking up preserves the structure
of the algebra but shrinks the scale of noncommutativity.
Written out completely, the counterterms
are
\bea \label{cntrtrms1}
\Delta \theta^{ij}(X)_{a,b} &=& \alpha^2 \sum_{c=1}^{N/2} \biggl[ 
\biggl( X^i_{2a-1,2c} - X^i_{2a-1,2c-1} \biggr) \cdot 
\biggl( X^j_{2c-1,2b-1} - X^j_{2c,2b-1} \biggr) - (i \leftrightarrow j)
\nonumber\\
&& {} \qquad + \biggl( X^i_{2a-1,2c} - X^i_{2a-1,2c-1} \biggr) \cdot 
\biggl( X^j_{2c-1,2b} - X^j_{2c,2b} \biggr) - (i \leftrightarrow j)
\nonumber\\
&& {} \qquad + \biggl( X^i_{2a,2c} - X^i_{2a,2c-1} \biggr) \cdot 
\biggl( X^j_{2c-1,2b-1} - X^j_{2c,2b-1} \biggr) - (i \leftrightarrow j)
\nonumber\\
&& {} \qquad + \biggl( X^i_{2a,2c} - X^i_{2a,2c-1} \biggr) \cdot 
\biggl( X^j_{2c-1,2b} - X^j_{2c,2b} \biggr) - (i \leftrightarrow j)
\biggr]
\eea
comprising difference terms in rows within blocks multiplying those in 
columns within blocks. This again gives rise to only nearest neighbour
interaction terms.

It is straightforward to check that the counterterms vanish for the case
of diagonal $X^i$, \ie\ $\theta^{ij} = 0$. Thus for the commutative 
case (see appendix~A), we do not need to perform any matrix 
``renormalization'' -- the commutative case is akin to a 
superrenormalizable field theory. For the noncommutative plane on the 
other hand, shifting away the diagonal matrices makes these 
counterterms vanish.
\\
With the above restrictions to $\{ O1 \}$ matrices, the $\Delta \theta$
counterterms simplify. For an $\{ O1 \}$ algebraic structure, nonzero 
terms arise only for $\Delta \theta^{ij}_{a,b}$ where 
$b=a,a\pm 1, a\pm 2$. Assuming (\ref{commlimitcondn1}) and 
(\ref{commlimitcondn2}) to hold, we can order the matrix elements so that
the diagonal elements are leading ($\{ O0 \}$) and the off-diagonal 
elements are subleading ($\{ O1 \}$). We can evaluate the above 
expression for the counterterms, rearranging and simplifying a bit. This 
gives, \eg,
\bea \label{cntrtrmsO1}
\Delta \theta^{ij}_{a,a+1} &=& \al^2 (X^i_{2a,2a+1}) \cdot \biggl[ 
( X^j_{2a,2a} - X^j_{2a-1,2a-1} ) + ( X^j_{2a+2,2a+2} - X^j_{2a+1,2a+1} )
\nonumber\\
{} && \qquad \qquad \ \ \ 
+ ( X^j_{2a-1,2a} - X^j_{2a,2a-1} ) - ( X^j_{2a+1,2a+2} - X^j_{2a+2,2a+1} )
\biggr] - (i \leftrightarrow j) \nonumber\\
{} && \sim \ {\cal O}(\{ O1 \}) [{\rm mass \ terms}] + 
{\cal O}(\{ O1 \}^2)
\eea
Define the change in noncommutativity under blocking up arising from 
such an infinitesimal noncommutative deformation as
\bea
{\delta \theta^{ij}_{ab} \over \delta X^k_{c,c+1}} \equiv 
{\biggl({\tilde \theta}^{ij}_{ab} - \theta^{ij}_{ab} \biggr) \over 
\delta X^k_{c,c+1}} = 
\ {\biggl( \alpha [\theta^{ij}]^{\rm blocked-up}_{ab} + \Delta 
\theta^{ij}_{ab} \biggr) - \theta^{ij}_{ab} \over \delta X^k_{c,c+1}}
\eea
Let us now focus for simplicity on purely diagonal $X^i$, \ie\ 
$\theta^{ij}=0$ (see appendix~A for the commutative case). Now turn on 
a small amount of off-diagonal component to the matrices $X^i$ so that 
$\theta^{ij}\neq 0$. Then 
\be
{\delta \theta^{ij}_{ab} \over \delta X^k_{c,c+1}}
= {\Delta \theta^{ij}_{ab} \over \delta X^k_{c,c+1}}
\ee
Focussing only on the leading order ${\cal O}(\{ O1 \})$ terms in
equations (\ref{cntrtrmsO1}), one can deduce that under blocking up 
\be\label{dtheta=0}
{\delta \theta^{ij}_{a,a+1} \over \delta X^k_{2c,2c+1}}
= \al^2 \biggl(
[( X^j_{2c,2c} - X^j_{2c-1,2c-1}) + ( X^j_{2c+2,2c+2} - X^j_{2c+1,2c+1})]
\delta^i_k - (i \leftrightarrow j)\biggr) \ \delta_{c,a},
\ee
Thus under blocking up, $\theta^{ij}$ is sensitive to differences in vevs
in nearest neighbour blocks as well (compare the changes in (\ref{theta0}) 
after turning on small bits of off-diagonal modes). 

The important lesson to learn from this technical subsection is that 
blocking up does not exhibit any ``chaotic'' behaviour. The change in 
noncommutativity under blocking up behaves in a seemingly controlled 
fashion -- only nearest neighbour blocks contribute to the way 
$\theta^{ij}$ changes. Thus assuming we have fixed gauge invariance and
assuming conditions (\ref{commlimitcondn1}) and (\ref{commlimitcondn2}) 
on the matrix elements (vevs) to hold, we see that blocking up appears to
change $\theta^{ij}$ in a not-too-jagged manner.

Focus now on the commutative limit. Assume that blocking up 
nearest-neighbour matrix elements is physically equivalent to averaging
over nearest neighbour D-branes, \ie\ nearest neighbour vevs are 
ordered appropriately as in the previous subsection (see Appendix A for 
some elaboration on this ordering in the commutative case and 
quasi-linear brane-chains). Then since the presence of a sufficiently
small averaging constant $\al<1$ scales down the change in 
noncommutativity (\ref{dtheta=0}) under blocking up, $\theta^{ij}$ 
appears to flow back to the commutative limit. In this case, 
$\theta^{ij}=0$ appears to be a $stable$ ``fixed point'' of this matrix
coarse-graining. This suggests that perhaps we can attribute 
meaning to notions such as renormalization group flows, relevance 
and irrelevance of operators, c-theorems, universality classes,
fixed points and other paraphernalia of Wilsonian renormalization in the
framework of D-brane geometries (\ref{branebgnd}) as well. Indeed, it 
would appear that D-brane configurations violating conditions 
(\ref{commlimitcondn1}) and (\ref{commlimitcondn2}) would generically 
exist, giving whole families of nontrivial algebras that might never 
exhibit shrinking off-diagonality under blocking up. Such algebras, in 
addition to those with infinite matrix correlation lengths, could be 
thought of as akin to exotic nontrivial fixed points. It would be very 
interesting to understand such nontrivial fixed points and indeed the 
structure of possible flows under this matrix coarse-graining.

\section{Conclusions and speculations}

In this work, we have studied D-brane configurations of the form 
(\ref{branebgnd}), where $\theta^{ij}(X)$ is in general spatially
varying. We have restricted attention to $\theta^{0i}=0$, \ie\ no 
timelike noncommutativity. Further we have made certain key assumptions 
involving gauge invariance. Given these assumptions, we have seen that 
blocking up D-brane configurations at the level of matrix variables 
shrinks off-diagonality in various classes of geometries, provided
certain conditions are satisfied by the matrices. Since the worldvolume
scalars describe the motion of the branes in the transverse space, this 
matrix coarse-graining would appear in a sense to be a coarse-graining
of the background spacetime in which the branes move. Further, blocking 
up seems to not do anything sporadic in such geometries. This suggests 
that there is sensible physics underlying such a matrix coarse-graining.

Besides the above, we have also implicitly assumed certain stability 
properties obeyed by (\ref{branebgnd}). The following example serves
to illustrate this point. Consider $N$ D0-branes in the presence of a
constant 4-form Ramond-Ramond field strength, as in the Myers effect
\cite{myers99}. The leading potential terms for the transverse scalars 
$X^i$ from the nonabelian Born-Infeld Lagrangian are of the form 
$\ {\cal L} \sim {\rm Tr} \biggl( -[X^i, X^j]^2 + 
C \epsilon^{ijk}X^iX^jX^k \biggr), \ i,j,k=1,2,3,\ $
where $W(X)=C\epsilon_{ijk}X^iX^jX^k$ arises from the Chern-Simons terms
coupling the D0-branes to the 4-form background field strength ($C$ is a
dimensionful constant). This gives an equation of motion schematically of
the form $\ \sum_m [X^m, [X^i, X^m]]-{\del W\over \del X^i}=0, \ $
which on substituting (\ref{branebgnd}) can be solved by $\ \theta^{ij}(X)
=[X^i, X^j] \sim C \epsilon^{ijk} X^k$. This solution, representing a 
fuzzy 2-sphere noncommutative geometry, turns out to have lower energy 
than the commutative solution with $X^i=0$ and is thus stable 
energetically. In principle, more general $W(X)$ can be used to solve 
for nontrivial field configurations, \ie\ nontrivial $\theta^{ij}(X)$. 
In general, different background fields in string theory would give rise
to different noncommutative backgrounds. We have assumed that D-brane 
configurations such as (\ref{branebgnd}) 
minimize energy given the background fields that have been turned on.
Thus we have restricted attention to static or quasi-static backgrounds
which can be treated as stable over some timescale characteristic to the
system -- small fluctuations have been assumed to not cause runaways 
from the background in question, at least on timescales long relative to 
some characteristic timescales intrinsic to the system. This assumption
of stability or meta-stability is a nontrivial one and corresponds to 
the assumptions of thermodynamic equilibrium or quasi-equilibrium in 
critical systems. Stability in D-brane systems might possibly (but not
necessarily) be enforced by supersymmetry.

\begin{figure}
\bc
\epsfig{file=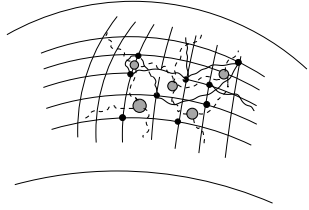, width=4cm}
\caption{$CARTOON$ : Pixellating spacetime with a fuzzy D-brane lattice 
: the dark dots are the locations of the microscopic D-branes in 
transverse space while the lighter dots represent block D-branes (the
worldvolume coordinates have been suppressed). The curvy lines are 
condensed open strings between the D-brane pixels.}
\label{fig3}
\ec
\end{figure}

If we restrict attention to geometries that admit a Riemannian limit,
one can make a few general arguments drawing intuition from the Riemannian
limit. One can make a normal coordinate expansion about any point on a
smooth manifold, expanding the metric as a flat piece with quadratic
corrections involving the curvature. One can then define notions such as
parallel transport to translate the tangent space at a given point to
neighbouring points. By analogy, consider schematically expanding 
(\ref{branebgnd}) as
\be
[X^i, X^j] = i\theta^{ij}(X) = i\theta^{ij}(x_0) + \ldots
\ee
$x_0$ is a ``point'' on the space.
Then the leading constant term on the right hand side looks like flat
noncommutative planes tiled together, with the $\ldots$ representing
curvature corrections to this noncommutative tangent space approximation.
Such an expansion can of course only be sensible in some sort of large
$N$ limit, where there exists a smooth Riemannian approximation. Even 
then, one needs to define a precise notion of a ``point'' in a 
noncommutative space to define the leading constant term and tangent
space approximation (see figure~\ref{fig3}).
More physically, defining a ``point'' precisely would necessarily be 
intertwined with formulating a precise definition of locality in a 
general noncommutative geometry.

Note that the tangent space approximation itself suggests that we have 
a constant $B$-field $B\sim{1\over \theta_0}$ turned on (as in the 
Seiberg-Witten limit \cite{sw9908}). The curvature corrections would
then suggest that if $H=dB=0$, \ie\ we have vanishing field strengths,
the limiting Riemannian manifold is flat. If on the other hand, 
$H=dB\neq 0$, it is curved. This is further corroborated by realizing
that vanishing string sigma model $\beta$-functions at leading order in
$\al'$ and $g_s$ give $\ R+H^2+\ldots\sim 0\ $.
This suggests that at least in some class of geometries, one can realize 
a Riemannian limit expanding about weak $H$-field strengths. $dB\neq 0$,
\ie\ nonzero curvature then suggests that there are sources for strings 
to end on, \ie\ D-branes, which serve to define pixellated 
approximations to the smooth geometry.
Blocking up the D-brane pixellations in the large $N$ limit then means
coarse-graining over the resolution of the pixellation. If one does not
probe too closely, one cannot resolve the individual pixels and a smooth
description emerges\footnote{Heuristically we expect that closed string
probes only see the smooth surface while open strings can probe the 
D-brane pixellation, as in, \eg, \cite{shenker95}, \cite{dkps}.}. 
In particular, since the tangent space
approximation consists of noncommutative planes where blocking up does 
seem to give a smooth limit, one would think blocking up would lead to
sensible physics for noncommutative planes patched together. However,
this chain of thoughts is rife with caveats. As in various lattice 
discretizations of smooth objects, it appears hard to define notions of 
topology here, for example the notion of topological proximity of points.
In fact, the noncommutative plane is itself only an ultralocal 
approximation and one needs to define even the notion of an open 
neighbourhood for a noncommutative geometry\footnote{For example, 
consider the spherical membrane \cite{kabattaylor} in Matrix theory, 
built out of D0-branes, \ie\ a fuzzy sphere.
Locally this is an object with D2-brane charge, but globally it has zero 
D2-brane charge. In the large $N$ limit, blocking up the representation
matrices of the $SU(2)$ algebra that builds up the fuzzy sphere does in
fact exhibit some self-similarity locally in the matrices, as we have 
seen. However, globally there might be subtleties having to do, for 
example, with parallel transport along the sphere at the level of matrix
variables.}. It would be interesting to understand how far block-brane 
techniques can be pushed towards the recovery of smooth manifold 
invariants along these lines. Thinking optimistically, maybe matrix 
scaling relations are all that are required to organize D-brane 
configurations, without really having to recover metrics.

At a worldsheet level, at each level of pixellation or block iteration 
order, one expects that the string sigma model is sensitive to the fact
that one has decimated some open string modes. In particular, the 
$B_{NSNS}$ term in the sigma model should reflect this decimation. It 
would be interesting to understand if this term and the way it flows
under blocking up can be organized in a sensible way, perhaps along the
lines of \cite{mukhi86}. To obtain an associative $*$-product for
nonconstant $\theta^{ij}$, one requires \cite{corschiap0101} that 
$\theta_{ij}^{-1}$ be closed, \ie\ $dB\equiv d(\theta^{-1})=0$ (see also
\cite{alekreckschom}). For the more general case where the algebra is 
not associative, it is tempting to guess that the open string sector 
still retains some remnant closed string (background) modes. In this 
case, one expects a noncommutative and nonassociative algebra of vertex
operators in general. Intuitively, blocking up and averaging over near 
neighbour open string modes would give a remnant closed string piece that
is now to be treated as part of the smooth background (no closed string
fluctuations). This new background now has new near neighbour open string
fluctuations which we then average over again, and so on ad infinitum. 
The new background is not necessarily described by an associative 
algebraic structure since we have absorbed some average closed string 
modes in its construction by blocking up. Since multiplication of 
$N\times N$ matrices is associative, one cannot hope to represent such
a nonassociative structure using them. It would be interesting to 
conjure up possible matrix-like structures that possess nonassociative 
multiplication, perhaps involving nonassociative $*$-products as in
\cite{corschiap0101}, where blocking up techniques as in this work may
be applied to gain insight into string algebras and spacetime geometry.
\vspace{6mm}

{\small 
{\bf Acknowledgments:}
Many of these ideas germinated and partially fructified towards the end of
my stay at Cornell -- I have benefitted greatly from discussions with
Philip Argyres, Swapneel Mahajan, Marco Moriconi, Vatche Sahakian and
Henry Tye on aspects of noncommutativity, as well as various condensed
matter folk there. Discussions pertinent to this paper with Philip
Argyres, Dave Morrison, Mukund Rangamani and Ashoke Sen have been very
useful. I have also benefitted from conversations with Paul Aspinwall,
Frederik Denef, Sergei Gukov, Tom Mehen, Ilarion Melnikov, Djordje Minic,
Horatiu Nastase, Gary Shiu and Mark Stern. Finally, it is a great pleasure
to thank Ronen Plesser for innumerable discussions, suggestions and
advice, without which this work would not have reached this shape.  I am
grateful to the members of the Theory Group at Newman Lab, Cornell, where
this work commenced, for the warm hospitality and congenial atmosphere
there. This work is partially supported by NSF grant DMS-0074072.
}

\appendix
\section{$\theta^{ij} = 0$ : the commutative case}

\begin{figure}
\bc
\epsfig{file=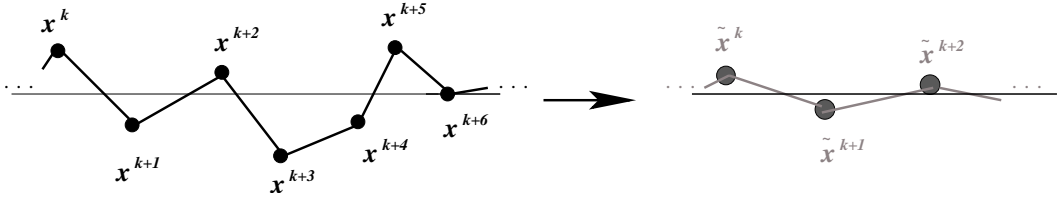, width=14cm}
\caption{A quasi-linear brane-chain. The dots are locations of D-branes
in transverse space, labelled before and after blocking up by $x^k$ and
${\tilde x}^k$ respectively (the worldvolume coordinates have been 
suppressed).}
\label{fig4}
\ec
\end{figure}

The commutative case seems harder in some ways. In fact overall blocking
up branes seems a priori artificial and ad hoc here, essentially because
points in the moduli space of a field theory define inequivalent vacua
of the system -- averaging over disjoint vacua does not seem a natural
thing to do drawing analogies with usual renormalization in field theory.
However, we shall formally block away and see where this takes us. It
turns out that there are parallels between what follows and usual
renormalization in field theory.\\
Consider $N$ D-branes with $[X^i, X^j] = \theta^{ij} = 0$. The $X^i$ are
scalars in the adjoint of $U(N)$. We further assume for now that there is
no additional matter so that we have sixteen supercharges. Then the $X^i$
can all be diagonalized and put in the form
\be \label{commutingbefore}
X^i = {\rm diag}[\ldots, \ x^i_k, \ x^i_{k+1}, \ x^i_{k+2}, \
x^i_{k+3}, \ \ldots]
\ee
The matrix element $X^i_{k,k} = x^i_k$ gives the position of the $k$-th
D-brane in the $i$-direction. Blocking up clearly preserves the diagonality
of the matrix, giving after the first iteration
\be \label{commutingafter}
{\tilde X}^i = {\rm diag}[\ldots, \ \al(x^i_k + x^i_{k+1}),\
\al(x^i_{k+2} + x^i_{k+3}), \ \al(x^i_{k+4} + x^i_{k+5}), \ \ldots]
\ee
This clearly preserves $\theta^{ij} = 0$, so that the algebra is preserved
in form. Thus there are no counterterms.
\\
However, it clearly reduces the rank of the matrix, going from $N$ branes 
to $N/2$ branes. In the large $N$ limit though, this of course preserves
rank. Indeed, for infinite $N$, the limit of the blocking up iterations
yields washed out D-branes, whose positions are averaged over those of 
the microscopic D-branes. Note that in carrying out this averaging, we 
have implicitly assumed that the matrix entries are ordered so that the 
nearest neighbour matrix entries indeed correspond to D-branes that are
physically close by -- this is necessary if blocking up is to sensibly 
represent physical averaging of the degrees of freedom. This ordering of
the matrix elements can of course be achieved for commuting matrices by
appropriate gauge transformations acting on the $X^i$, in all cases where
the configuration picks out one spatial dimension as dominant. For example,
if the D-branes are arranged in an approximately linear chain (see \eg\
figure~\ref{fig4}), blocking up will make physical sense as a spatial 
averaging of D-branes.\\
Now, assuming such a spatial ordering of the matrix entries, it is 
reasonable to demand that blocking up should yield a D-brane whose position
is averaged over the positions of the constituent D-branes in the block. 
Then it is clear that we must fix $\al={1\over 2}$ as the uniform 
averaging constant. This then means that the new block D-brane is located
at the center of mass of the original D-branes.\\
Instead of averaging over $2\times 2$ blocks, let us average over 
$3\times 3$ blocks. Then this clearly still preserves the form of the
algebra and yields washed out D-branes with each iteration, as before. 
Averaging over $k\times k$ blocks is equivalent to averaging over the
spatial locations of $k$ D-branes so that we must fix $\al={1\over k}$.
Note that this is identical to the noncommutative plane, where the 
existence of the translation isometries fixed $\al={1\over k}$ for 
$k\times k$ block averaging.

Consider again the D-brane configuration before blocking 
(\ref{commutingbefore}) and the corresponding configuration after
blocking up (\ref{commutingafter}), with $\alpha=1/2$. For the moment,
let us restrict attention to the case with $16$ supercharges and further
to the generic point in the moduli space where the gauge group is 
completely broken to $U(1)^\infty$, \ie\ $x_i\neq x_j$. Then the 
masses of the lightest half-BPS states represented by strings stretched
between, for \eg, branes $1,2,3,4$ before blocking up are proportional to
$|x_i - x_j|, \ i\neq j$. The corresponding mass of the lightest half-BPS
state between branes $\{ 12 \}$ and $\{ 34 \}$ after blocking up is 
\bea\label{massineq}
m_{(12),(34)}=
\biggl| {\vx_1+\vx_2 \over 2} - {\vx_3+\vx_4 \over 2} \biggr|
&=& \biggl| {1\over 4} (\vx_1-\vx_3 + \vx_1-\vx_4 + \vx_2-\vx_3 + 
\vx_2-\vx_4) \biggr| \nonumber\\
&\leq& {1\over 4} \biggl( |\vx_1-\vx_3| + |\vx_1-\vx_4| + 
|\vx_2-\vx_3| + |\vx_2-\vx_4| \biggr)
\eea
Thus the mass of the lightest (half-BPS) open string modes stretched 
between the block branes is less than the average of the masses of the 
open string modes stretched between the original branes. After blocking up,
one cannot distinguish between the $\{ 13 \}, \{ 14 \}, \{ 23 \}, \{ 24 \}$
strings -- they all get mapped to the single string stretched between the
block branes $\{ 12 \}$ and $\{ 34 \}$. String web states that stretch
between, for example, branes $\{ 123 \}$ and other combinations of branes
outside the appropriate curves of marginal stability, also get mapped under
blocking up to the single string state stretched between the block branes.\\
Let us now construct a low energy effective
field theory keeping all massless modes and the lightest massive modes -- 
this is to be treated in a Wilsonian sense as an effective theory good for 
studying processes only upto energy scales less than the masses of the 
open string states that we have averaged over.
Such a low energy effective theory (including the lightest massive modes)
about the blocked configuration is thus less sensitive, on the average, to
high energy processes than the original configuration (at a generic vacuum).
This holds only on the average -- the inequality only holds for the average
of the short distance open string states, not individually.
We can continue this process of averaging over short distance open string
modes, defining block D-branes iteratively, focussing only on the nearest
neighbour open string modes in each iteration. With each such iteration,
the nearest neighbour modes will be most dominant in determining the
effective interactions in the theory.
Blocking up thus appears in some sense to induce a ``flow'' in moduli space
towards lower energies.

The above inequality (\ref{massineq}) assumes a generic vacuum -- 
collinear branes give an equality in the above equation. This suggests 
that thinking of D-branes as lattice points might only make sense in some
regions of moduli space. To give more perspective on this, let us return
to the ordering of D-brane arrangements at the level of matrix variables.
For a quasi-linear brane-chain, blocking up would retain quasi-linearity. 
Thus for such locations in moduli space, blocking up matrix elements 
would faithfully represent blocking up D-branes in physical space. This 
assumes spatial ordering of the $\vx_k$s (see figure~\ref{fig4})
where $\vx_k=(x_k^1,x_k^2, \ldots)\ $ is the vector representing the
location of the $k$-th D-brane in the transverse space. However, there
are hoards of gauge transformations that correspond to this 
same brane-chain in physical space. A gauge-invariant way to block up
would be to perform a weighted blocking up of each set of matrices in a 
given gauge orbit. A schematic example of such a weight is $\Pi_{i,j}\
e^{-|\vx_i-\vx_j|^2}$. This ensures that gauge equivalent configurations
with large differences in the eigenvalues $x_i$ are suppressed while 
matrices whose eigenvalue entries are near each other dominate, while
still retaining manifest gauge invariance.\\
In general, mapping a general spatial lattice-like structure to an ordering
of D-brane matrix eigenvalues that is faithful physically under blocking
up appears contrived\footnote{However it is not too hard to show that 
such mappings exist. Consider for example, an infinite 2-dimensional 
regular lattice of D-branes in physical 2-space. Then the matrices 
$X^1,X^2$ are both 2-way infinite. One can order the matrix eigenvalues,
ordering the negative eigenvalues by half-spiralling inwards to the 
origin in physical space and the positive eigenvalues by half-spiralling
outwards.}.
Indeed such mappings are perhaps only possible in some regions of 
moduli space\footnote{This might not be too surprising since identifying
D-branes with lattice sites that mock up a deconstructed dimension also
only works in some regimes of moduli space \cite{deconstopol}.}. It would
be interesting to study specific brane configurations, quivers and 
deconstructions to examine whether blocking up branes gives any new 
insights into the commutative case.

\section{Other ways to block up branes}

It is important to note that one can cook up other ways to block up 
D-branes. For example, consider blocking up as
\bea
X^i&=&{\rm diag}\biggl( x^i_1, x^i_2, x^i_3, x^i_4, x^i_5, x^i_6, 
\ldots \biggr)
\ \ra \ 
{\tilde X}^i={\rm diag}\biggl({x^i_1+x^i_2\over 2}{\bf 1_2}, 
{x^i_3+x^i_4\over 2}{\bf 1_2},{x^i_5+x^i_6\over 2}{\bf 1_2}, \ldots 
\biggr) \nonumber\\
{} && \ \ra \ {\tilde
{\tilde X}}^i={\rm diag}\biggl({x^i_1+x^i_2+x^i_3+x^i_4\over 2}{\bf 1_4},
{x^i_5+x^i_6+x^i_7+x^i_8\over 2}{\bf 1_4}, \ldots \biggr) \ \ra \ \ldots
\eea
This clearly does not reduce the rank of the gauge group even for finite
$N$. What we have done is to in fact enhance gauge symmetry, from a 
$U(k)\times U(k)$ in each block to a $U(2k)$. Indeed the limit of this 
kind of blocking up for $N$ branes is a $U(N)$ gauge theory. Thus we have
integrated $in$ the massive open string modes that were stretched between
noncoincident branes within a block. This kind of blocking up therefore
moves towards the ultraviolet of the field theory, as opposed to the 
infrared as before. \\
Indeed, consider the supergravity solution dual to this (\Nf) field theory
configuration on the Coulomb branch -- this is a multicenter D3-brane 
supergravity solution \cite{trivedi9811}. Then the centers in the 
corresponding block go from two separated $N$ brane centers to a single
$2N$ brane center. Physically what this means is that we have 
approximated a multicenter background by an averaged single
center background with the same mass -- to leading order, one does not 
resolve the separation between the centers. Thus this is blocking up in 
$spacetime$, while the previously discussed blocking up is in field theory
space\footnote{Trying to apply this spacetime blocking up to the 
noncommutative plane does not seem to lead to anything recognizable however
-- in fact, a bit of algebra seems to suggest that field theory blocking 
up seems to be the more appropriate thing to do there.}.

The basic point we are trying to make here is that since 
D-branes admit two dual descriptions, via gauge theory and via gravity,
we can can approximate D-brane systems by blocking them up in two 
manifestly different ways. What we have illustrated in the bulk of this
work is blocking up in field theory. This section gives a brief glimpse
of what blocking up in spacetime is like, as an approximation scheme.
Perhaps there are yet other physically relevant ways to block up branes
with new physics.

\end{document}